\documentclass[journal]{IEEEtran}

\usepackage{cite}
\usepackage[linkcolor=green, anchorcolor=blue]{hyperref}

\ifCLASSINFOpdf
  \usepackage[pdftex]{graphicx}
\else
  \usepackage[dvips]{graphicx}
\fi

\usepackage{amsmath}
\usepackage{amssymb}
\usepackage{algorithm}
\usepackage{algorithmic}
\usepackage{url}
\usepackage{color}
\usepackage{makecell}

\ifCLASSOPTIONcompsoc
 \usepackage[caption=false,font=normalsize,labelfont=sf,textfont=sf]{subfig}
\else
 \usepackage[caption=false,font=footnotesize]{subfig}
\fi

\usepackage{booktabs}

\begin{document}

\title{Scalable Multi-task Edge Sensing via Task-oriented Joint Information Gathering and Broadcast}

\author{Huawei~Hou,~\IEEEmembership{Student Member,~IEEE,}
        Suzhi~Bi,~\IEEEmembership{Senior Member,~IEEE,}
        Xian~Li,~\IEEEmembership{Senior Member,~IEEE,}
        Shuoyao~Wang,~\IEEEmembership{Senior Member,~IEEE,}
        Liping~Qian,~\IEEEmembership{Senior Member,~IEEE,}
        and Zhi~Quan,~\IEEEmembership{Senior Member,~IEEE}
\thanks{An earlier version of this paper was presented in part at the 13th IEEE/CIC International Conference on Communications in China (ICCC) 2024 \cite{ICCC}.}
\thanks{H.~Hou, S.~Bi, X.~Li, S.~Wang and Z.~Quan are with the College
of Electronics and Information Engineering, Shenzhen University, Shenzhen,
China 518060 (e-mail: 2350432016@email.szu.edu.cn, \{bsz,  xianli, sywang, zquan\}@szu.edu.cn).}
\thanks{L.~Qian is with the College of Information Engineering, Zhejiang University of Technology, Hangzhou, Zhejiang, China, 310023. (e-mail: lpqian@zjut.edu.cn)}

}

\maketitle

\begin{abstract}
The recent advance of edge computing technology enables significant sensing performance improvement of Internet of Things (IoT) networks. In particular, an edge server (ES) is responsible for gathering sensing data from distributed sensing devices, and immediately executing different sensing tasks to accommodate the heterogeneous service demands of mobile users. However, as the number of users surges and the sensing tasks become increasingly compute-intensive, the huge amount of computation workloads and data transmissions may overwhelm the edge system of limited resources. Accordingly, we propose in this paper a scalable edge sensing framework for multi-task execution, in the sense that the computation workload and communication overhead of the ES do not increase with the number of downstream users or tasks. By exploiting the task-relevant correlations, the proposed scheme implements a unified encoder at the ES, which produces a common low-dimensional message from the sensing data and broadcasts it to all users to execute their individual tasks. 
To achieve high sensing accuracy, we extend the well-known information bottleneck theory to a multi-task scenario to jointly optimize the information gathering and broadcast processes. 
We also develop an efficient two-step training procedure to optimize the parameters of the neural network-based codecs deployed in the edge sensing system. 
Experiment results show that the proposed scheme significantly outperforms the considered representative benchmark methods in multi-task inference accuracy. Besides, the proposed scheme is scalable to the network size, which maintains almost constant computation delay with less than 1\% degradation of inference performance when the user number increases by four times. 
\end{abstract}


\IEEEpeerreviewmaketitle
\section{Introduction}
\subsection{Motivations and Contributions}
\begin{figure}[!t]
\centering
\includegraphics[width=3.3in]{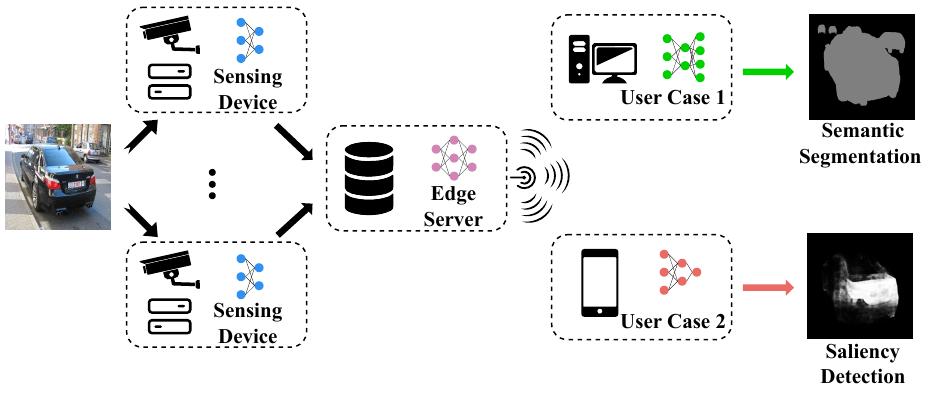}
\caption{A schematic of the considered wireless edge sensing system.
}
\label{fig_sim}
\end{figure}

\IEEEPARstart{I}{n} the field of Internet of Things (IoT), edge computing contributes significantly in reducing latency and enhancing the digital experiences of diverse time-sensitive applications, such as virtual reality (VR), Internet of Vehicles (IoV), industrial internet, and Metaverse \cite{ec1, ec2, ec3, EL_survay,ISCC,bi}. 
As shown in Fig. \ref{fig_sim}, an IoT edge sensing system typically includes upstream edge sensing devices (SDs), an edge server (ES), and downstream users. 
In particular, the ES plays the central role in gathering the sensing information from the upstream SDs \cite{ec4} (e.g., the monitoring photos of vehicles) and then disseminating to downstream users for performing different computing tasks (e.g., semantic segmentation and saliency detection). The goal of the edge sensing system is to ensure highly efficient execution of the users' sensing tasks, such as high inference accuracy and low end-to-end sensing latency. 
However, as we move towards future intelligent IoT networks, the dramatic growth of data volume and more complex computing workload poses significant challenges to the performance of edge sensing systems with limited communication bandwidth and computing power \cite{IOT_band}. 
Extensive studies have been dedicated to optimizing user-server allocation \cite{ec5}, computation offloading \cite{IOT2}, and service caching \cite{ec6} to enhance the utilization of communication and computing resources, which, however, cannot fundamentally solve the resource scarcity problem as they fail to effectively control the sensing workloads admitted to the edge computing system. The scalability of edge sensing performance against workload increase remains a challenging problem.

The recent advance of \emph{semantic communication} \cite{survey} technology offers a promising solution to building a scalable edge sensing system. Instead of transmitting bulky raw data, it transmits only minimal task-relevant semantic features extracted from raw data. Essentially, semantic communication relies on neural networks (NNs) to perform source coding in a task-oriented manner. 
Besides, NN-based joint source-channel coding (JSCC) techniques are devised to combat the distortion of noisy wireless channels, which have demonstrated excellent task execution performance, communication efficiency, and noise robustness under different application scenarios \cite{ec8, ec7, deepJSCC, sc4}. 
The application of semantic communication in edge sensing systems can effectively reduce the overhead in both upstream and downstream communications. 
Especially, it is sufficient for the ES to broadcast a common semantic feature to enable efficient multi-task executions of the users, which is more bandwidth-efficient compared to transmitting dedicated message to each user. \cite{ei2, b3}. 
Following a task-oriented design principle to maximize the sensing accuracy, the upstream data gathering and the downstream information dissemination processes are inter-dependent and thus should be jointly optimized in an end-to-end manner. One major difficulty is to establish a holistic analytical framework for designing the NN-based encoders/decoders to fully exploit the shared semantic features and simultaneously combat the concatenated noisy channels in the upstream and downstream processes. Besides, end-to-end training of NNs can be costly in practice, often requiring excessively long training time and a vast amount of training data.  

To tackle the above challenges, we propose in this paper a scalable semantic edge sensing framework for multi-task execution, as shown in Fig. \ref{fig_sim}, which jointly optimizes the upstream information gathering and downstream broadcast. 
In particular, we provide information-theoretical analysis, efficient training algorithm, and hardware implementation to support the design of the proposed edge sensing scheme. Our detailed contributions are summarized as follows: 

\begin{itemize}

    	\item  \textcolor{black}{   \textbf{Scalable multi-task edge sensing framework:} The proposed scheme exploits the task-relevant correlations in both the upstream data gathering and downstream information dissemination. Facing the dissimilar service requirements of multiple inference tasks, the proposed method gathers only one copy of sensing data from the upstream SDs and broadcasts a common message to users. As such, the system is \textit{scalable} to the network size, as the communication overhead and the computation workload of the ES does not increase with the \textit{number of users or inference tasks.}}
    
	    \item  \textcolor{black}{   \textbf{Joint optimization of information gathering and broadcast:} To capitalize on the semantic correlation in the upstream and downstream data processing, we jointly optimize the information gathering and broadcast to achieve an optimum end-to-end inference performance. Specifically, we propose an efficient two-step training procedure to optimize the NN-based codecs using a multi-task information bottleneck (IB) method.} 

        \item \textbf{Channel-adaptive Codec Design:} We take different approaches to optimize the NN-based codecs to deal with the variation of wireless channel conditions in the upstream and downstream. Considering the stable channel conditions between the fixed SDs and ES, we propose a channel-aware pooling (CAP) method based on average SNR to adjust the dimension of the transmitted feature of the SDs to balance the consumed bandwidth and robustness against noise. Meanwhile, the downstream channels are highly dynamic and heterogeneous due to users' mobility. In this case, we use the Monte Carlo sampling method to simulate a wide range of SNR variations in the training process.  

\end{itemize}

We evaluate the performance of the proposed edge sensing scheme using both simulations and hardware implementations. Experiment results show that the proposed scheme significantly outperforms representative benchmark methods in multi-task inference performance (e.g., 10.1\% better than the downlink unicast scheme for saliency detection task). Meanwhile, it is scalable to the increase of network size, which maintains almost constant computation delay with less than 1\% performance degradation when the user number increases by four times. In comparison to the unicast benchmark method, not only its sensing performance drops by as large as $20\%$ but also suffers a $4$ times increase of the edge encoding delay.

\begin{figure*}[!t]
\centering
\includegraphics[width=6in]{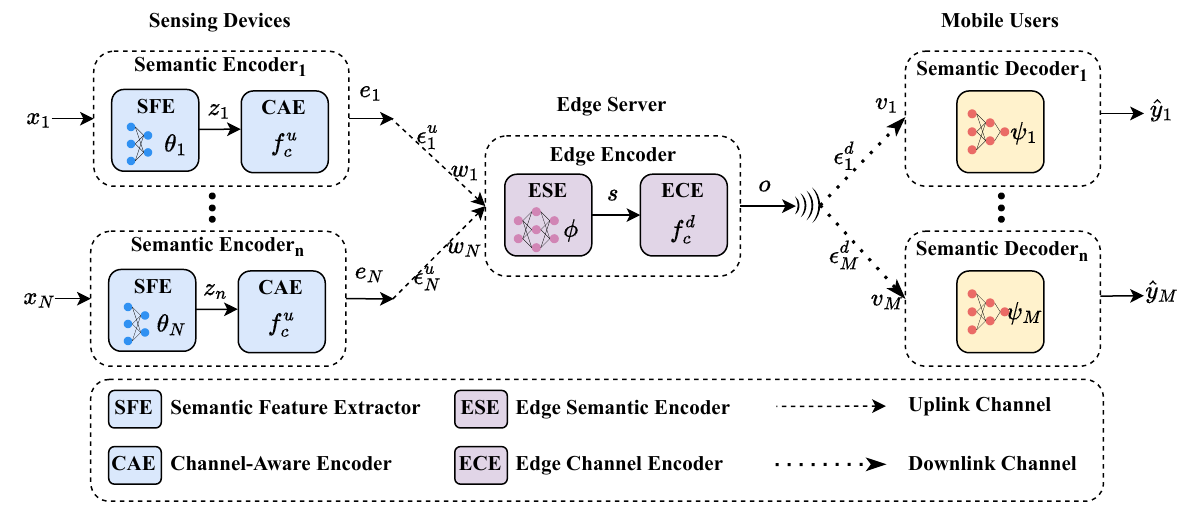}
\caption{Block diagram of the proposed scalable edge sensing system.}
\label{fig_2}
\end{figure*}

\subsection{Related Works} \label{related}
The recent development of semantic communication \cite{survey} has made significant strides in improving the performance of edge sensing systems for IoT applications. 

\subsubsection{Semantic Information Gathering}
Instead of transmitting raw sensory data, edge sensing systems allow the SDs 
to transmit semantic features (usually extracted by NNs) to the ES. Then, with the noisy features received, the ES performs specific inference tasks, e.g., image classification or objective detection. 
Specifically, Shao et al. \cite{shao1} developed an edge inference system, where a mobile device extracts and transmits the task-relevant features to an ES to finish task inference. The system is designed with the IB theory \cite{IB} to achieve a tradeoff between the transmitted information and the inference accuracy. Shao et al. \cite{shao2} further extended the system to a distributed scenario with the distributed IB \cite{DIB}, where multiple edge devices capture a sensing object and transmit the extracted features to the ES to conduct cooperative inference. Inspired by multi-task learning, Sheng et al. \cite{Sheng_multi} proposed a multi-task semantic communication system for natural language processing (NLP), where a transmitter extracts features with a common text encoder, and transmits to a receiver, which uses different tasks' decoder for inference. In addition to the single-modal source described above, Zhang et al. \cite{zhang_multi} proposed a unified deep learning-enabled semantic communication system (U-DeepSC), which can handle three modal information sources simultaneously and use a unified decoder at the receiver to execute downstream tasks.

\subsubsection{Semantic Information Dissemination}
Different from semantic information gathering, semantic sensing information dissemination refers to a transmitter (e.g., SD or ES) conveying the extracted semantic features of sensing data to multiple destined users, which then use their own decoders to complete inference task. 
Compared to the relatively stable uplink channel, i.e., from static SDs to the ES, the downlink channels are often more dynamic and complex due to the mobility of users. For improving channel robustness, Ding et al. \cite{ei2} designed a multi-user image transmission system, which trains the NN-based image codec by sampling varying the channel conditions of different users. Wu et al. \cite{b3} considered the scenario of large channel differences between users and incorporated a semantic feature weight adjustment mechanism in the design of a semantic broadcast system adaptive to different users' channels. 
Besides the complex channels, heterogeneous downstream tasks also pose a challenge to efficient semantic information dissemination. To address this challenge, Gong et al. \cite{unicast} proposed a multiple-encoder-multiple-decoder (MEMD) system, where a transmitter uses a series of channel encoders to encode the semantic features and transmit to the corresponding downstream users, i.e., unicast method, which requires a separate bandwidth allocation for each user. To further improve bandwidth efficiency, Yang et al. \cite{b6} proposed a task-oriented broadcast framework for edge inference, where an edge transmitter extracts and broadcasts features to multiple edge devices to conduct multi-task inference.

The above-mentioned works on semantic information gathering and dissemination encompass one-to-one, many-to-one, or one-to-many scenarios, wherein the semantic information is transmitted either on the uplink or downlink direction. 
In this paper, as shown in Fig. \ref{fig_sim}, we consider a many-to-many scenario where multiple upstream SDs transmit sensing information to an ES for the multiple downstream users to complete task inference. 
Meanwhile, a joint design is required to harness the task-relevant correlation in the uplink and downlink transmissions and to tackle the concatenated noisy channels. 
Hence, in this paper, we establish a holistic analytical framework for designing the NN-based encoders/decoders of both the upstream and downstream and design an efficient two-step training strategy for the NN-based codecs.

\subsection{Organizations}
The rest of the paper is structured as follows. In Section \ref{problem}, we provide an overview of the system model and outline the design objectives. Section \ref{method} delves into the technical intricacies of the proposed framework, detailing the design of upstream and downstream codecs. In Section \ref{result}, we evaluate the performance of the proposed scheme and compare it with representative benchmarks. The paper is finally concluded in Section \ref{clu}. 

\subsection{Notations}
In this paper, we denote upper-base letters (such as $X, Y, Z$) and lower-base letters (such as $\boldsymbol{x, y, z}$) as random variables and corresponding realizations, respectively. The entropy of $Y$ is denoted as $H(Y)$, and the mutual information between $X$ and $Y$ is denoted as $I(X; Y)$. The Kullback-Leibler (KL) divergence between two probability distributions $p(\boldsymbol{x})$ and $q(\boldsymbol{x})$ is denoted as $D_{KL}(p \parallel q)$.

\section{System Model and Problem Formulation} \label{problem}
The schematic of the proposed multi-task edge sensing system is illustrated in Fig. \ref{fig_2}. Specifically, we consider $N$ static SDs collaboratively monitoring a target object, e.g., cameras taking multi-view photos of a car in Fig. \ref{fig_sim}.   
Each SD utilizes its local NN to extract the semantic features from the raw sensing data, performs JSCC, and then transmits the encoded features to the ES through a noisy uplink wireless channel. Here, we consider that the transmissions of the $N$ SDs are in orthogonal channels without co-channel interference.  
After receiving all the $N$ features, the ES re-encodes them into a 
common feature and broadcasts to $M$ downstream mobile users. Upon receiving the common feature, the $M$ users perform different but correlated sensing tasks to obtain the inference results. For example, some users perform semantic segmentation task while some others perform saliency detection task using the received common feature.

\subsection{System Model}
We denote the observed data samples of the $N$ SDs as $\boldsymbol{x}=\{\boldsymbol{x}_1, \boldsymbol{x}_2, ..., \boldsymbol{x}_N\}$. The ground-truth labels of the inference tasks at the $M$ users are denoted as $\boldsymbol{y}=\{\boldsymbol{y}_1, \boldsymbol{y}_2, ..., \boldsymbol{y}_M\}$.

\subsubsection{Semantic Encoders at SDs}
To extract task-relevant features from observations $\boldsymbol{x}$, we deploy a series of semantic feature extractor (SFE) $ f_{\boldsymbol{\theta}} = \{ f_{\boldsymbol{\theta}_1}, f_{\boldsymbol{\theta}_2}, ..., f_{\boldsymbol{\theta}_N} \}$ parameterized by $\boldsymbol{\theta} = \{  \boldsymbol{\theta}_1, \boldsymbol{\theta}_2, ..., \boldsymbol{\theta}_N \}$ on the SDs. 
Accordingly, the upstream feature extraction of data samples sensed by SD $n$ is represented as
\begin{equation}
    \boldsymbol{z}_n = f_{\boldsymbol{\theta}_n}(\boldsymbol{x}_n), \ \ n=1, 2, ..., N.
\end{equation}
Before transmission, we adopt a non-learnable channel-aware encoder (CAE) $f_c^u$ to further compress the extracted features and normalize the transmission power to unity, where the output is a $K^u$-dimension vector denoted as
\begin{equation}
    \boldsymbol{e}_n = f_{c}^{u}(\boldsymbol{z}_n) , \ \ n=1, 2, ..., N, 
\end{equation}
with $||\boldsymbol{e}_n||_2^2$ = 1. We denote the compressed information of SDs as $\boldsymbol{e}=\{\boldsymbol{e}_1, \boldsymbol{e}_2, ..., \boldsymbol{e}_N\}$, which is then transmitted to the ES through uplink wireless channels. 
     \textcolor{black}{In this paper, we consider block fading uplink channels, and the received signal at the ES is
	\begin{equation} \label{Z_to_W}
		\boldsymbol{w}_n = h_n \cdot \boldsymbol{e}_n+\boldsymbol{\epsilon}_n^u ,\ \ n=1, 2, ..., N. 
	\end{equation}
	Here, $h_n$ denotes the channel coefficient between the $n$th SD and the ES, which is assumed i.i.d. among the $N$ SDs. For simplicity of illustration, we assume without loss of generality that the channel has unit power ($\mathbb{E}[|h_n|^2] = 1$), and distinguish channel conditions by setting different noise powers. For a general Rician fading channel, $h_n$ follows a complex Gaussian distribution $\mathcal{CN}(\sqrt{a_n/(a_n+1)}, 1/(a_n+1))$ with parameter $a_n\geq 0$. Notice that by setting $a_n = 0$, it reduces to a Rayleigh fading channel. Besides, 
	$\boldsymbol{\epsilon}_n^u \sim \mathcal{CN}(0, \sigma_n^2\boldsymbol{I})$ denotes a receiver complex Gaussian noise. 
	We define the average SNR of the $n$th uplink channel as 
	\begin{equation}
		\bar{\Upsilon}_n^u = \frac{\mathbb{E}[|h_n|^2]}{\sigma_n^2},\ \ n=1, 2, ..., N. \tag{4}
	\end{equation}
}

\subsubsection{Edge Encoder at ES}
The ES collects all the received noisy features denoted by $\boldsymbol{w} = \{ \boldsymbol{w}_1, \boldsymbol{w}_2, ..., \boldsymbol{w}_N  \}$, then re-encodes the received features into a common feature for downstream broadcast. 
We denote the edge semantic encoder as $f_{\boldsymbol{\phi}}$ parameterized by $\boldsymbol{\phi}$. The encoding process of the edge semantic encoder (ESE) is
\begin{equation}
    \boldsymbol{s}=f_{\boldsymbol{\phi}}(\boldsymbol{w}). 
\end{equation}
Before broadcast, we use a non-learnable edge channel encoder (ECE) $f_c^d$ to normalize the transmission power of the re-encoded common feature, where the normalized output is a $K^d$-dimension vector
\begin{equation}
    \boldsymbol{o}=f_{c}^{d}(\boldsymbol{s}), 
\end{equation}
with $||\boldsymbol{o}||_2^2$ = 1. 
     \textcolor{black}{Then, the ES broadcasts the common feature $\boldsymbol{o}$ through downlink channels. Accordingly, the $m$th user receives
	\begin{equation} \label{o_to_v}
		\boldsymbol{v}_m= g_m \cdot \boldsymbol{o}+\boldsymbol{\epsilon}_m^d,\ \  m=1, 2, ..., M,  \tag{7}
	\end{equation}
	where $g_m$ is the downlink fading channel coefficient. Similar to the uplink channel in (\ref{Z_to_W}), $g_m$ has unit power and is modeled as a Rician fading channel with parameter $b_m\geq 0$. 
	$\boldsymbol{\epsilon}_m^d \sim \mathcal{CN}(0, \sigma_m^2\boldsymbol{I}) $ is a complex Gaussian noise. Accordingly, we define the average SNR of the $m$th downlink channel as 
	\begin{equation}
		\bar{\Upsilon}_m^d = \frac{\mathbb{E}[|g_m|^2]}{\sigma_m^2},\ \  m=1, 2, ..., M.  \tag{8}
\end{equation}}
\subsubsection{Decoders at Users}
After receiving $\boldsymbol{v}_m$, the $m$th user decodes to generate a predicted output. With the decoder $f_{\boldsymbol{\psi}}=\{f_{\boldsymbol{\psi}_1}, f_{\boldsymbol{\psi}_2}, ..., f_{\boldsymbol{\psi}_M} \}$ parameterized by $\{\boldsymbol{\psi}_1, \boldsymbol{\psi}_2, ..., \boldsymbol{\psi}_M\}$, the prediction of the $m$th user is denoted as
\begin{equation}
    \hat{\boldsymbol{y}}_m = f_{\boldsymbol{\psi}_m}(\boldsymbol{v}_m),\ \  m=1, 2, ..., M.
\end{equation}

The overall process of upstream information gathering, edge processing, and downstream broadcast and inference can be modeled as a Markov chain:
\begin{equation} \label{Markov}
    Y \rightarrow X \stackrel{\boldsymbol{\theta}}\rightarrow  Z \rightarrow W \stackrel{\boldsymbol{\phi}}\rightarrow  S  \rightarrow V \stackrel{\boldsymbol{\psi}}\rightarrow \hat{Y}.
\end{equation}

\subsection{Problem Formulation}
In the considered edge sensing system, we need to optimize the parameters of the encoders and decoders, i.e., $\{\boldsymbol{\theta}, \boldsymbol{\phi}, \boldsymbol{\psi}\}$, to maximize the inference accuracy. For simplicity of analysis, we simplify the complete Markow chain in (\ref{Markov}) to $Y \rightarrow X  \rightarrow W  \rightarrow V$. 
To balance between data fit and generalization, we follow the information theoretical design principle IB to seek a sufficient and minimal representation of the observed data $\boldsymbol{x}$. Specifically, we optimize the encoders at the SDs ($\boldsymbol{\theta}$) and the ES ($\boldsymbol{\phi}$) by maximizing the following IB expression
\begin{equation} \label{IB1}
    \max_{\boldsymbol{\theta} , \boldsymbol{\phi} }~ I_{\boldsymbol{\theta} , \boldsymbol{\phi}}(Y;V)-\beta \cdot I_{\boldsymbol{\theta} , \boldsymbol{\phi}}(X;V).
\end{equation}
The first mutual information item in (\ref{IB1}) is to maximally reserve task-relevant information in features received by downstream users. The second item is to minimize sample correlation between raw data and the extracted features for better generalization performance. $\beta$ is a non-negative hyper-parameter to control the trade-off between the two items. The design of the decoder $\boldsymbol{\psi}$ depends on the nature of heterogeneous downstream task. It will be jointly optimized with the edge encoder as an auxiliary task detailed in Section \ref{encoder_decoder}. 

Directly computing the mutual information terms in (\ref{IB1}) involves non-tractable integration over empirical probability distributions such as $p(\boldsymbol{x})$ and $p(\boldsymbol{v}|\boldsymbol{x})$. In practice, we can apply variational IB methods that use NNs to approximate the probability distribution and solve the minimization problem in (\ref{IB1}) with numerical Monte Carlo sampling and end-to-end training \cite{shao1}. However, the naive end-to-end training method may involve in high training complexity, low sample efficiency, and poor convergence performance. Hence, in this paper, we propose a reduced-complexity joint optimization method leveraging the task-relevant correlation between the upstream and downstream processes.

\section{Task-oriented Joint Information Gathering and Broadcast} \label{method}
In this section, we propose a two-step method that sequentially optimizes the sensor-side semantic encoder $\boldsymbol{\theta}$ and the edge encoder $\boldsymbol{\phi}$. The user-side decoders $\boldsymbol{\psi}$ are simultaneously designed and optimized with the edge encoder $\boldsymbol{\phi}$ as auxiliary tasks' decoders.

\subsection{Design of Upstream Encoder} \label{object_en}
We first consider the Markov chain $Y \rightarrow X \rightarrow Z \rightarrow W$ in the upstream to design the semantic encoder $\boldsymbol{\theta}$ at the sensor side. 
To conserve task-relevant information in the upstream data gathering stage, we take a task-oriented design perspective and suppose that the $M$ downstream tasks are executed at the ES after receiving $\boldsymbol{w}$ from the sensors. 
Following the IB design principle, we optimize the encoder parameters $\boldsymbol{\theta}$ by solving the following multi-task information bottleneck (MIB) minimization problem \cite{MIB}
\begin{equation} \label{IB_fe}
    \min_{\boldsymbol{\theta}}~ -\sum_{m=1}^{M} \alpha _m I_{\boldsymbol{\theta} }(Y_{m};W)+\beta \cdot \sum_{n=1}^{N}I_{\boldsymbol{\theta} _n}(X_n;Z_n).
\end{equation} 
Here, the first item corresponds to the mutual information of downstream tasks, where $\alpha _m$'s are the weighting parameters for different tasks. The second item corresponds to the minimum representation objective for better generalization performance. Notice that the CAE is a non-learnable function determined by the average uplink SNR, we therefore apply the IB principle between $X_n$ and $Z_n$ to guide the parameter training of SFE. The detailed design is shown in Section \ref{imb}. Then, to compute the MIB in (\ref{IB_fe}), we expand the mutual information expression to minimize the following multi-task IB objective after neglecting some constant terms (see Appendix \ref{P_VI} for detailed derivations)
\begin{align} \label{L_IB}
    & \mathcal{L}_{MIB}(\boldsymbol{\theta}) \notag \\ & \ \ = \mathbf{E}_{p(\boldsymbol{x}, \boldsymbol{y})} \Bigg  \{ \mathbf{E}_{p_{\boldsymbol{\theta}}(\boldsymbol{z}| \boldsymbol{x})} \Bigg [   -\sum_{m=1}^{M} \alpha_m \log p (\boldsymbol{y}_{m}|\boldsymbol{w}) \Bigg ] \notag \\
    & \quad + \beta \cdot \sum_{n=1}^{N} D_{KL}(p_{\boldsymbol{\theta}_n}(\boldsymbol{z}_n| \boldsymbol{x}_n) \parallel  p(\boldsymbol{z}_n) ) \Bigg  \}. 
\end{align}
Computing $\mathcal{L}_{MIB}$ involves high-dimensional integrals over general distributions, e.g., $p(\boldsymbol{z}_n)$ and $p (\boldsymbol{y}_{m}|\boldsymbol{w})$. 
Instead, we adopt the variational information bottleneck (VIB) \cite{deepIB} method to approximate the intractable integrals in (\ref{L_IB}). 
Specifically, we introduce an auxiliary NN parameterized by $\boldsymbol{\lambda}$, whose input is the received features $\boldsymbol{w}$ of the ES in (\ref{Z_to_W}) and the output is the predicted distribution of label $\boldsymbol{y}_{m}$, denoted by $q_{\boldsymbol{\lambda}} (\boldsymbol{y}_{m}|\boldsymbol{w}),\  m= 1,\cdots, M $. We use $q_{\boldsymbol{\lambda}} (\boldsymbol{y}_{m}|\boldsymbol{w})$ to approximate $p (\boldsymbol{y}_{m}|\boldsymbol{w})$. Besides, we introduce a normal distribution $q(\boldsymbol{z}_n) \sim N(0, \boldsymbol{I})$ to approximate $p(\boldsymbol{z}_n), \ n=1,\cdots,N$. 
Accordingly, we rewrite (\ref{L_IB}) as the following multi-task variational IB
\begin{align} \label{L_VIB1}
    &  \mathcal{L}_{VMIB}(\boldsymbol{\theta}, \boldsymbol{\lambda}) \notag \\ & \ \ \ = \mathbf{E}_{p(\boldsymbol{x}, \boldsymbol{y})} \Bigg \{ \mathbf{E}_{p_{\boldsymbol{\theta}}(\boldsymbol{z}| \boldsymbol{x})} \Bigg [   -\sum_{m=1}^{M} \alpha_m \log q_{\boldsymbol{\lambda}} (\boldsymbol{y}_{m}|\boldsymbol{w}) \Bigg ] \notag \\
    &  \ \ \ + \beta \cdot \sum_{n=1}^{N} D_{KL}(p_{\boldsymbol{\theta}_n}(\boldsymbol{z}_n| \boldsymbol{x}_n) \parallel  q(\boldsymbol{z}_n) ) \Bigg \},  
\end{align}
which provides an upper bound of $\mathcal{L}_{MIB}$. Therefore, the MIB minimization problem in (\ref{IB_fe}) is transformed to minimizing an upper bound $\mathcal{L}_{VMIB}$ given in (\ref{L_VIB1}). The detailed derivation of (\ref{L_IB}) and (\ref{L_VIB1}) is shown in Appendix \ref{P_VI}.

\subsection{Implementation of the Upstream Encoder} \label{imb}

In this subsection, we describe the implementation details of minimizing (\ref{L_VIB1}).

\begin{figure}[!t]
\centering
\includegraphics[width=3.5in]{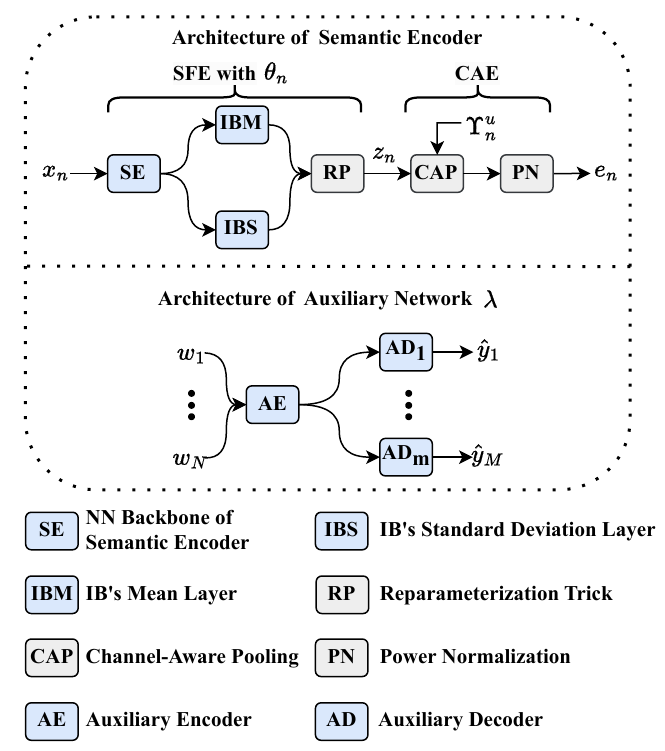}
\caption{The architecture of the upstream semantic encoder. The light blue modules in the figure represent the parameter modules that need to be trained, and the gray modules are the functional modules that do not need to be trained.}
\label{fig_3}
\end{figure}

\subsubsection{Network Architecture}
The architecture of the semantic encoder deployed on the $n$th edge SD is shown in Fig. \ref{fig_3}.
With an observation data input $\boldsymbol{x}_n$, we first use a NN backbone to extract semantic features and compress data, i.e., the SFE module in Fig. \ref{fig_3}. 
Here, we use two NN layers (the IBM and IBS modules in Fig. \ref{fig_3}) to obtain the mean and standard deviation of distribution $p_{\boldsymbol{\theta}_n}(\boldsymbol{z}_n| \boldsymbol{x}_n)$.
Then, we apply the reparameterization trick for the forward propagation of the data, which is detailed later in (\ref{reparameter}). 
After obtaining the $\boldsymbol{z}_n$ from SFE, we then use a CAE module to further process the $\boldsymbol{z}_n$ to output the transmitted signal $\boldsymbol{e}_n$. Specifically, we use a channel-aware pooling (CAP) module to adjust the output dimension based on uplink channel SNR. Intuitively, we apply a more refined pooling to produce a larger output dimension, and vice versa. 
After power normalization, the encoded features $\boldsymbol{e}_n$ will be transmitted to the ES through the uplink channel. 

\subsubsection{Channel-Aware Pooling}  \label{CAP}
Using a larger dimension of the output feature can effectively enhance the robustness of edge inference against channel noise, however, also consumes more transmission bandwidth. To achieve a balanced performance between the two competing objectives, we propose a CAP method to flexibly adjust the output dimension of $\boldsymbol{e}_n$ based on channel conditions. 
As shown in Fig. \ref{fig_3}, the output from NN of the edge device is processed by a tunable \emph{adaptive-pooling} layer. The pooling output size $l_p$ is decided by the average uplink channel SNRs. A larger SNR leads to a larger pooling output size $l_p$, which translates to a smaller output dimension of $\boldsymbol{e}_n$. In practice, the uplink SNRs between the static SDs and the ES are stable and indeed similar. We therefore decide the pooling output size based on the mean of the average SNR of the $N$ uplink channels, which is a fixed value for a specific deployment scenario. With the proposed CAP, the semantic encoder can improve bandwidth utilization by adjusting the output dimension of the model under different SNR conditions instead of redesigning the entire model structure.
After pooling and power normalization (PN), the processed features are transmitted to the ES. The ES will upsample the received signal to restore a fixed-size feature, i.e., the same size as before pooling. 

\subsubsection{Training Method} \label{LF_SE}
Here, we apply the Monte Carlo sampling method to optimize the objective function in (\ref{L_VIB1}) with respect to \{$\boldsymbol{\theta}, \boldsymbol{\lambda}$\}. 
We first focus on the first term $-\sum_{m=1}^{M} \alpha_m \log q_{\boldsymbol{\lambda}} (\boldsymbol{y}_{m}|\boldsymbol{w})$. The goal is to find the best parameters of $\boldsymbol{\lambda}$ to maximize the corresponding distribution $q_{\boldsymbol{\lambda}} (\boldsymbol{y}_{m}|\boldsymbol{w})$ across all training samples. According to the \emph{negative log-likelihood criterion} \cite{ud}, we have 
\begin{align} \label{ls}
     \hat{\boldsymbol{\lambda}} &= \mathop{\arg}\max_{\boldsymbol{\lambda}} \left[ \sum_{b=1}^{B}  \sum_{m=1}^{M} \alpha_m  \log q_{\boldsymbol{\lambda} } (\boldsymbol{y}_{m, b}|\boldsymbol{w}_b) \right] \notag \\
     &= \mathop{\arg}\min_{\boldsymbol{\lambda}} \left[ -  \sum_{b=1}^{B} \sum_{m=1}^{M} \alpha_m  \log q_{\boldsymbol{\lambda} } (\boldsymbol{y}_{m, b}|\boldsymbol{w}_b) \right] \notag \\ 
     &= \mathop{\arg}\min_{\boldsymbol{\lambda}} \left[ \sum_{b=1}^{B} \sum_{m=1}^{M}\alpha_m  l_{m}(\boldsymbol{y}_{m, b}, \hat{\boldsymbol{y}}_{m, b}(\boldsymbol{w}_b;\boldsymbol{\lambda })) \right], 
\end{align}
where $B$ is the number of the training samples, and $l_{m}(\cdot , \cdot )$ represents the $m$th loss function (depending on the specific task, e.g., $l_m=-  \sum_{b=1}^{B} \mathrm{softmax}_{\boldsymbol{y}_b} f (\boldsymbol{w}_b;\boldsymbol{\lambda}) $ for the multi-classification cross-entropy loss \cite{ud}) to compute the difference between the ground-truth label $\boldsymbol{y}_{m}$ and the predicted output $\hat{\boldsymbol{y}}_{m}(\boldsymbol{w};\boldsymbol{\lambda})$. 

Then, the distribution $p_{\boldsymbol{\theta}_n}(\boldsymbol{z}_n| \boldsymbol{x}_n)$ is commonly parameterized by utilizing multivariate Gaussian distribution \cite{shao2, b6},  $\mathcal{N}(\boldsymbol{z}_n| f_{\boldsymbol{\theta}_n}^{\boldsymbol{\mu}}(\boldsymbol{x}_n) , f_{\boldsymbol{\theta}_n}^{\boldsymbol{\sigma}}(\boldsymbol{x}_n))$. Notice that $f_{\boldsymbol{\theta}_n}^{\boldsymbol{\mu}}(\boldsymbol{x}_n)$ and $f_{\boldsymbol{\theta}_n}^{\boldsymbol{\sigma}}(\boldsymbol{x}_n)$ are $K^z$-dimensional mean vector and standard deviation vector determined by $\boldsymbol{\theta}_n$, which are the output of the IBM and IBS module in Fig. \ref{fig_3}. That is $f_{\boldsymbol{\theta}_n}^{\boldsymbol{\mu}}(\boldsymbol{x}_n) = (\mu_{n, 1}, ..., \mu_{n, K^z}) \in \mathbb{R}^{K^z} $,  $f_{\boldsymbol{\theta}_n}^{\boldsymbol{\sigma}}(\boldsymbol{x}_n) = (\sigma_{n, 1}, ..., \sigma_{n, K^z}) \in \mathbb{R}^{K^z}$.  
After obtaining $f_{\boldsymbol{\theta}_n}^{\boldsymbol{\mu}}(\boldsymbol{x}_n)$ and $f_{\boldsymbol{\theta}_n}^{\boldsymbol{\sigma}}(\boldsymbol{x}_n)$, we use the reparameterization trick \cite{reparameterization} to compute
\begin{equation} \label{reparameter}
    \boldsymbol{z}_n = f_{\boldsymbol{\theta}_n}^{\boldsymbol{\mu}}(\boldsymbol{x}_n) + f_{\boldsymbol{\theta}_n}^{\boldsymbol{\sigma}}(\boldsymbol{x}_n) \odot \boldsymbol{\epsilon}^z, 
\end{equation}
where $\odot$ denotes Hadamard product, and $\boldsymbol{\epsilon}^z$ is a standard normal distribution, i.e., $\boldsymbol{\epsilon}^z \sim \mathcal{N}(0, \boldsymbol{I}) $. 
Meanwhile, the distribution $q(\boldsymbol{z}_n)$ is treated as a centered isotropic standard multivariate Gaussian distribution \cite{shao2}. Hence, we formulate the KL-divergence in (\ref{L_VIB1}) as
\begin{equation} \label{kl}
\begin{split}
  & \  D_{KL}(p_{\boldsymbol{\theta}_n}(\boldsymbol{z}_n| \boldsymbol{x}_n) \parallel  q(\boldsymbol{z}_n) ) \\
  & \ \    = \sum_{k=1}^{K^z} \left \{  \frac{\mu_{n, k}^2 + \sigma_{n, k}^2 -1 }{2} - \log \sigma_{n, k} \right \}.
\end{split}
\end{equation}
Finally, with (\ref{ls}) and (\ref{kl}), we rewrite (\ref{L_VIB1}) as the following expression
\begin{align} \label{loss_VIB_MC}
& \mathcal{L}_{VMIB}(\boldsymbol{\theta}, \boldsymbol{\lambda}) \notag \\ & \ \ \simeq \frac{1}{B} \sum_{b=1}^{B}  \Bigg \{ \sum_{m=1}^{M}  \alpha_m  l_{m}(\boldsymbol{y}_{m, b}, \hat{\boldsymbol{y}}_{m, b}(\boldsymbol{w}_b;\boldsymbol{\lambda })) \notag \\
& \ \ + \beta \sum_{n=1}^{N} \sum_{k=1}^{K^z} \left \{  \frac{\mu_{n, k, b}^2 + \sigma_{n, k, b}^2 -1 }{2} - \log \sigma_{n, k, b} \right \} \Bigg \} ,  
\end{align}
with the training dataset $\{\boldsymbol{x}_{1, b}, ..., \boldsymbol{x}_{N, b}, \boldsymbol{y}_{1, b}, ..., \boldsymbol{y}_{M, b}\}_{b=1}^{B}$. During training, $\boldsymbol{\theta}$ and $\boldsymbol{\lambda}$ in (\ref{loss_VIB_MC}) can be optimized by minimizing $\mathcal{L}_{VMIB}$ with Stochastic Gradient Descent (SGD) based backpropogation. 
The training procedures are shown in Algorithm \ref{alg_feature extractor}. After training, we fix the parameters of the semantic encoder $\boldsymbol{\theta}$ and discard the auxiliary NN $\boldsymbol{\lambda}$. 

\begin{algorithm}[!t] 
    \caption{ Training Procedures of Semantic Encoder } 
    \label{alg_feature extractor} 
    \renewcommand{\algorithmicrequire}{\textbf{Input:}}
    \renewcommand{\algorithmicensure}{\textbf{Output:}}
    \begin{algorithmic}[1]
		\REQUIRE ~~\\ 
		Training set $\mathcal{D}^{train}$. 
		\ENSURE ~~\\ 
		The optimized semantic encoder $\boldsymbol{\theta}$.\\
		\STATE  Initialize parameters of $\boldsymbol{\theta}$ and the auxiliary NN $\boldsymbol{\lambda}$.
        \REPEAT
        \STATE Sample a mini-batch $\{\boldsymbol{x}_{1, b}, ..., \boldsymbol{x}_{N, b}, \boldsymbol{y}_{1, b}, ..., \boldsymbol{y}_{M, b}\}_{b=1}^{B_c}$ from $\mathcal{D}^{train}$ randomly;
        \STATE Compute instantaneous SNRs $\{\Upsilon_{1, b}^u, ...,\Upsilon_{N, b}^u\}_{b=1}^{B_c}$ of the uplink channels.
        \FOR{$b=1$ to $B_c$}
        \FOR{$n=1$ to $N$}
        \STATE Extract semantic features $\boldsymbol{z}_{n,b}$ with SFE; 
        \STATE Produce the output $\boldsymbol{e}_{n,b}$ based on $\Upsilon_{n,b}^u$; 
        \STATE Transmit $\boldsymbol{e}_{n,b}$ to the ES.
        \ENDFOR
        \STATE Gather noisy features $\boldsymbol{w}_{b} = \{ \boldsymbol{w}_{1,b}, \boldsymbol{w}_{2,b}, ..., \boldsymbol{w}_{N,b}  \}$.
        \FOR{$m=1$ to $M$}
        \STATE Compute $\hat{\boldsymbol{y}}_{m, b}(\boldsymbol{w}_b;\boldsymbol{\lambda })$ with $\boldsymbol{\lambda}$.
        \ENDFOR
        \ENDFOR
        \STATE Update parameters of $\boldsymbol{\theta}$ and $\boldsymbol{\lambda}$ by minimizing (\ref{loss_VIB_MC}).
        \UNTIL Convergence condition is met.	
    \end{algorithmic}
\end{algorithm}

\begin{figure}[!t]
\centering
\includegraphics[width=3.5in]{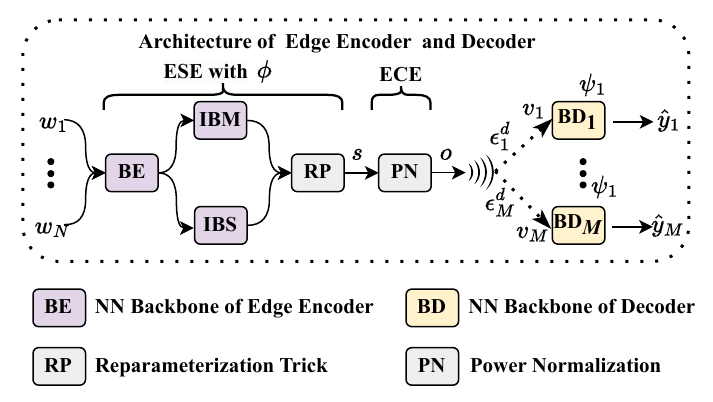}
\caption{The architecture of the edge encoder and decoder. The mauve and yellowish modules are the parameter modules that need to be trained for the edge encoder and decoder, respectively.} 
\label{fig_4}
\end{figure}

\subsection{Implementation of the Edge Encoder and Downstream Decoder} \label{encoder_decoder}
As shown in Fig. \ref{fig_4}, with the received noisy features $\boldsymbol{w}$ from the trained semantic encoder, we now focus on the downstream semantic feature re-encoding and decoding process in Markov chain $W \rightarrow S \rightarrow V \rightarrow \hat{Y}$, and optimize the NN-based edge encoder and downstream user-side decoder. 

\subsubsection{Objective Function}
To achieve a trade-off between downstream tasks' inference performance and feature broadcast overhead, we optimize the parameters $\boldsymbol{\phi}$ of the edge encoder by solving the following MIB problem 
\begin{equation} \label{IB_de}
    \min_{\boldsymbol{\phi}}~ -\sum_{m=1}^{M} \gamma _m I_{\boldsymbol{\phi} }(Y_{m};V_{m})+ \eta \cdot I_{\boldsymbol{\phi}}(S;W),
\end{equation}
where the first item is the mutual information of downstream tasks, $\gamma _m $'s are the weighting parameters for different tasks, and the second item corresponds to the minimum representation objective for better generalization performance. $\eta$ is a non-negative hyper-parameter to control the trade-off between the two items. 

Following the similar approach as designing the upstream encoder, it is equivalent to solve (\ref{IB_de}) by minimizing the following objective
\begin{align} \label{L_IB_de}
  &  \mathcal{L}_{MIB}(\boldsymbol{\phi}) \notag \\ & \ \ = \mathbf{E}_{p(\boldsymbol{w}, \boldsymbol{y})} \Bigg \{ \mathbf{E}_{p_{\boldsymbol{\phi}}(\boldsymbol{s}|\boldsymbol{w})} \Bigg[ -\sum_{m=1}^{M} \gamma_m \log p (\boldsymbol{y}_{m}|\boldsymbol{v}_{m}) \Bigg] \notag \\ 
    & \ \ \quad + \eta \cdot D_{KL}(p_{\boldsymbol{\phi}}(\boldsymbol{s}|\boldsymbol{w}) \parallel p(\boldsymbol{s})) \Bigg \}. 
\end{align}
Then, we adopt the VIB method to approximate the intractable integrals in (\ref{L_IB_de}). Specifically, we apply a series of decoder parameterized by $\boldsymbol{\psi}=\{ \boldsymbol{\psi}_1, \boldsymbol{\psi}_2, ..., \boldsymbol{\psi}_M \}$ to generate distribution $q_{\boldsymbol{\psi}_m } (\boldsymbol{y}_{m}|\boldsymbol{v}_{m})$ to approximate $p (\boldsymbol{y}_{m}|\boldsymbol{v}_{m})$. Besides, we introduce a normal distribution $q(\boldsymbol{s}) \sim N(0, \boldsymbol{I})$ to approximate $p(\boldsymbol{s})$. 
Hence, we can rewrite (\ref{L_IB_de}) as
\begin{align}
\label{L_VIB_de}
&   \mathcal{L}_{VMIB}(\boldsymbol{\phi}, \boldsymbol{\psi}) \notag  \\ 
& \ \  =\mathbf{E}_{p(\boldsymbol{w}, \boldsymbol{y})} \Bigg \{ \mathbf{E}_{p_{\boldsymbol{\phi}}(\boldsymbol{s}| \boldsymbol{w})}  \Bigg [   -\sum_{m=1}^{M} \gamma_m \log q_{\boldsymbol{\psi}_m } (\boldsymbol{y}_{m}|\boldsymbol{v}_{m}) \Bigg ] \notag  \\ 
& \quad \quad \quad +  \eta \cdot D_{KL}(p_{\boldsymbol{\phi}}(\boldsymbol{s}| \boldsymbol{w}) \parallel  q(\boldsymbol{s})) \Bigg \}.  
\end{align}
The detailed derivation is shown in Appendix \ref{P_VI_D}.

\subsubsection{NN Architecture of Edge Encoder and User Decoder}
The NN architecture of the edge encoder and decoder is shown in Fig. \ref{fig_4}. Specifically, the edge encoder uses a NN backbone to re-encode features, i.e., the ESE in Fig. \ref{fig_4}. The IBM module and the IBS module are used to generate the predicted mean and standard deviation of a variational probability distribution, which is similar in Section \ref{imb}. By applying the reparameterization trick, we obtain the re-encoded features $\boldsymbol{s}$.
Then, in this paper, we use a power normalization function as the ECE to normalize the $\boldsymbol{s}$ to the broadcast feature $\boldsymbol{o}$.

\subsubsection{Training Method} \label{LF_DE}
We use the Monte Carlo sampling method to optimize the objective function in (\ref{L_VIB_de}) with respect to \{$\boldsymbol{\phi}$, $\boldsymbol{\psi}$\}. For the first term $-\sum_{m=1}^{M} \gamma_m \log q_{\boldsymbol{\psi}_m } (\boldsymbol{y}_{m}|\boldsymbol{v}_{m}) $, we have
\begin{align} \label{ld}
     \hat{\boldsymbol{\psi}} &= \mathop{\arg}\max_{\boldsymbol{\psi}} \left[ \sum_{b=1}^{B}  \sum_{m=1}^{M} \gamma_m  \log q_{\boldsymbol{\psi}_m } (\boldsymbol{y}_{m,b}|\boldsymbol{v}_{m,b}) \right] \notag \\
     &= \mathop{\arg}\min_{\boldsymbol{\psi}} \left[ -   \sum_{b=1}^{B} \sum_{m=1}^{M} \gamma_m  \log q_{\boldsymbol{\psi}_m } (\boldsymbol{y}_{m,b}|\boldsymbol{v}_{m,b}) \right] \notag \\
     &= \mathop{\arg}\min_{\boldsymbol{\psi}} \left[ \sum_{b=1}^{B} \sum_{m=1}^{M}\gamma_m  l_{m}(\boldsymbol{y}_{m,b}, \hat{\boldsymbol{y}}_{m}(\boldsymbol{v}_{m,b};\boldsymbol{\psi }_{m})) \right].
\end{align}

\begin{algorithm}[!t] 
    \caption{ Training Procedures of Edge encoder and User Decoder } 
    \label{alg_decoder} 
    \renewcommand{\algorithmicrequire}{\textbf{Input:}}
    \renewcommand{\algorithmicensure}{\textbf{Output:}}
    \begin{algorithmic}[1]
		\REQUIRE ~~\\ 
		Training set $\mathcal{D}^{train}$, pre-trained semantic encoder $\boldsymbol{\theta}$. 
		\ENSURE ~~\\ 
		The optimized edge encoder $\boldsymbol{\phi}$ and user decoder $\boldsymbol{\psi}$.\\
		\STATE  Initialize model parameters of $\boldsymbol{\phi}$, $\boldsymbol{\psi}$.
        \REPEAT
        \STATE Sample a mini-batch $\{\boldsymbol{x}_{1, b}, ..., \boldsymbol{x}_{N, b}, \boldsymbol{y}_{1, b}, ..., \boldsymbol{y}_{M, b}\}_{b=1}^{B_c}$ from $\mathcal{D}^{train}$ randomly.
        \FOR{$b=1$ to $B_c$}
        \FOR{$n=1$ to $N$}
        \STATE Compute  semantic features $\boldsymbol{e}_{n,b}$ with the pre-trained semantic encoder and then transmit $\boldsymbol{e}_{n,b}$.
        \ENDFOR
        \STATE Gather noisy features $\boldsymbol{w}_b = \{ \boldsymbol{w}_{1,b}, \boldsymbol{w}_{2,b}, ..., \boldsymbol{w}_{N,b}  \}$;
        \STATE Re-encode $\boldsymbol{w}_b$ to $\boldsymbol{s}_b$ with ESE;
        \STATE Normalize power $\boldsymbol{s}_b$ to $\boldsymbol{o}_b$ and broadcast $\boldsymbol{o}_b$. 
        \FOR{$m=1$ to $M$}
        \STATE Receive $\boldsymbol{v}_{m,b}$ and compute $\hat{\boldsymbol{y}}_{m,b}(\boldsymbol{v}_{m,b};\boldsymbol{\psi }_m)$.
        \ENDFOR
        \ENDFOR
        \STATE Update parameters of $\boldsymbol{\phi}$ and $\boldsymbol{\psi}$ by minimizing (\ref{loss_VIB_de_MC}).
        \UNTIL Convergence condition is met.	
    \end{algorithmic}
\end{algorithm}

Then, we use the multivariate Gaussian distribution $ \mathcal{N}(\boldsymbol{s}| f_{\boldsymbol{\phi}}^{\boldsymbol{\mu}}(\boldsymbol{w}) , f_{\boldsymbol{\phi}}^{\boldsymbol{\sigma}}(\boldsymbol{w}))$ to parameterize $p_{\boldsymbol{\phi}}(\boldsymbol{s}| \boldsymbol{w}) $, where $f_{\boldsymbol{\phi}}^{\boldsymbol{\mu}}(\boldsymbol{w})$ and $f_{\boldsymbol{\phi}}^{\boldsymbol{\sigma}}(\boldsymbol{w})$ are $K^s$-dimensional mean vector and standard deviation vector determined by $\boldsymbol{\phi}$.
Specifically, they are the outputs of the IBM and IBS modules in Fig. \ref{fig_4}. That is, $f_{\boldsymbol{\phi}}^{\boldsymbol{\mu}}(\boldsymbol{w}) = (\mu_{s, 1}, ..., \mu_{s, K^s}) \in \mathbb{R}^{K^s} $,  $f_{\boldsymbol{\phi}}^{\boldsymbol{\sigma}}(\boldsymbol{w}) = (\sigma_{s, 1}, ..., \sigma_{s, K^s}) \in \mathbb{R}^{K^s}$. With the reparameterization trick, we have
\begin{equation} \label{reparameter_d}
    \boldsymbol{s} = f_{\boldsymbol{\phi}}^{\boldsymbol{\mu}}(\boldsymbol{w}) + f_{\boldsymbol{\phi}}^{\boldsymbol{\sigma}}(\boldsymbol{w}) \odot \boldsymbol{\epsilon}^s, 
\end{equation}
where $\boldsymbol{\epsilon}^s$ is a standard normal distribution, i.e., $\boldsymbol{\epsilon}^s \sim \mathcal{N}(0, \boldsymbol{I}) $. By treating the distribution $q(\boldsymbol{s})$ as a centered isotropic standard multivariate Gaussian distribution, we formulate the KL-divergence in (\ref{L_VIB_de}) as 
\begin{align} \label{kl_d}
      & \  D_{KL}(p_{\boldsymbol{\phi}}(\boldsymbol{s}| \boldsymbol{w}) \parallel  q(\boldsymbol{s})) \notag \\
      & \ \  = \sum_{k=1}^{K^s} \left \{  \frac{\mu_{s, k}^2 + \sigma_{s, k}^2 -1 }{2} - \log \sigma_{s, k} \right \}.
\end{align}
Finally, with (\ref{ld}) and (\ref{kl_d}), we rewrite (\ref{L_IB_de}) as 
\begin{align} \label{loss_VIB_de_MC}
& \mathcal{L}_{VMIB}(\boldsymbol{\phi}, \boldsymbol{\psi}) \notag \\
& \ \ \simeq  \frac{1}{B} \sum_{b=1}^{B}  \Bigg \{ \sum_{m=1}^{M} \gamma_m  l_{m}(\boldsymbol{y}_{m, b}, \hat{\boldsymbol{y}}_{m, b}(\boldsymbol{v}_{m, b};\boldsymbol{\psi }_m)) \notag  \\
& \ \ +  \eta \cdot \sum_{k=1}^{K^s} \left \{  \frac{\mu_{s, k, b}^2 + \sigma_{s, k, b}^2 -1 }{2} - \log \sigma_{s, k, b} \right \} \Bigg \}, 
\end{align}
with the training dataset $\{\boldsymbol{x}_{1, b}, ..., \boldsymbol{x}_{N, b}, \boldsymbol{y}_{1, b}, ..., \boldsymbol{y}_{M, b}\}_{b=1}^{B}$. 
The detailed training procedures of $\boldsymbol{\phi}$ and $\boldsymbol{\psi}$ are shown in Algorithm \ref{alg_decoder}.

\section{Numerical Results} \label{result}
In this section, we evaluate the performance of the proposed multi-task edge sensing method and compare it with representative benchmarks. 

\subsection{Experiment Setup}

\subsubsection{Dataset}
To establish a multi-task edge processing environment, we leverage the PASCAL-Context dataset \cite{Pascal}, i.e., a widely used multi-task dataset for computer vision research \cite{M3VIT}. It comprises a total of 10,103 RGB images, divided into 4,998 training images and 5,105 testing images. Each image is associated with multiple labels corresponding to different tasks. In this work, we consider 3 downstream users performing 3 different tasks: semantic segmentation, human parts segmentation, and saliency detection. To imitate the multi-view image capture scenario, we assume that each image is captured by four edge SDs, where each SD acquires a partial view of the image. Specifically, we reshape each RGB image in the dataset to a size of $3\times512\times512$ pixels and subsequently split it into four overlapping parts, from top-left to bottom-right, as shown in Fig. \ref{fig_8}. Each part has a dimension of $3\times384\times384$ pixels and is captured by one of the four edge SDs.

\subsubsection{Hardware}
We train our models offline on a high-performance server equipped with two Intel Xeon Silver 4210R CPUs, 256GB of memory, and two NVIDIA GeForce RTX 3090 GPUs, each with 24GB of video memory. After training, we deploy the models on resource-constrained edge devices to examine the inference performance. 
Specifically, we use 7 NVIDIA Jetson Nano as the 4 sensing devices and the 3 user devices, where each Jetson Nano has a Quad-core ARM Cortex-A57 MPCore processor, 4GB of memory, and a 128-core NVIDIA Maxwell architecture GPU. The detailed specifications of the Jetson Nano and Jetson Xavier platforms can be found on the official NVIDIA website\footnote{\url{https://developer.nvidia.com/embedded/jetson-modules}}.
We use a single NVIDIA Jetson Xavier as the ES, which has a 6-core NVIDIA Carmel Arm v8.2 CPU, 4GB of memory, and a 384-core NVIDIA Volta architecture GPU. 
It is worth noting that we aim to validate the real-time inference latency on resource-limit devices. The features are sent through Wi-Fi connected wireless devices, and we use simulated noise added to the received symbols to emulate channel fading and receiver noise. 

\begin{figure}[!t]
\centering
\includegraphics[width=3.5in]{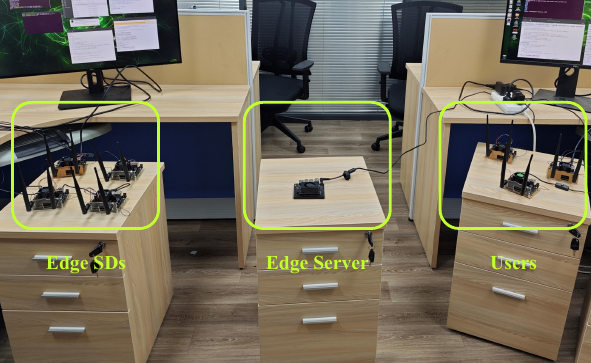}
\caption{ The edge sensing platform established with NVIDIA Jetson devices.}
\label{fig_j}
\end{figure}

\subsubsection{Channel Condition} 
\textcolor{black}{
	We consider that line-of-sight link exists for the uplink channels between fixed SDs and the ES, where we set $a_n =1$ for $h_n$ in (\ref{Z_to_W}). For simplicity of illustration, we consider a set of average uplink SNR $\bar{\Upsilon}^u \in \{-10, -5, 0, 5, 10\}$ dB for performance evaluation. Meanwhile, for the downlink channels between the ES and mobile users, we model them as Rayleigh fading channels without a strong line-of-sight link (setting $b_m=0$ in (\ref{o_to_v})) considering the users' mobility in a scattering environment. Then, each point in the following figures is an average performance of $20$ independent channel realizations.
}

\subsubsection{NN Architecture}
The hyper-parameters of the NN model used by the proposed method are shown in Table \ref{tab_network_p}. We use Resnet18 and DeepLabHead\footnote{\url{https://github.com/pytorch/vision/blob/main/torchvision/models/segmentation/deeplabv3.py}} (a well-known pixel-level image processing toolbox) as the backbone networks. Here, all Resnet18 backbones are pre-trained with the ImageNet dataset. 
Because we consider the adaptive output dimension of the CAP layer based on the uplink channel SNR, we use ``$c_1$" to denote the variable output dimension, which is equal to the pooling output size $l_p$. Specifically, we set the $l_p$ as $\{8, 7, 6, 5, 4\}$ for $\bar{\Upsilon}^u = \{-10, -5, 0, 5, 10\}$ dB, respectively. Besides, we use ``$c_2$" in the decoder to denote the output dimension of different sensing tasks, e.g., $21$ for image semantic segmentation, $7$ for human parts segmentation, and $1$ for saliency detection. \textcolor{black}{Notice that the output dimension of the SDs is much smaller than the original images ($2-7\%$ under different SNRs). Therefore, the required communication bandwidth of the proposed system is much smaller than that based on conventional bit-oriented communications.}

During training, we use the pixel-level cross-entropy loss function for semantic segmentation and human parts segmentation, and the pixel-level binary cross-entropy loss function for saliency detection. 
We use an Adam optimizer with a learning rate of $2\times10^{-4}$ and a weight decay of $1\times10^{-4}$. The batch size is 8 and the training epochs is 100. The weighting parameters $\alpha_m$ and $\gamma_m$ are both set to 1.0, 2.0, and 5.0 for semantic segmentation, human parts segmentation, and saliency detection, respectively, and the trade-off parameters $\beta$ and $\eta$ are both  $10^{-7}$. Besides, the same settings are used for the training of \{$\boldsymbol{\theta}$, $\boldsymbol{\lambda}$\} and \{$\boldsymbol{\phi}$, $\boldsymbol{\psi}$\}. 

\begin{table}[!t] 
\centering
\caption{Architecture of NN in Proposed Method}
\resizebox{\columnwidth}{!}{%
\begin{tabular}{lll} 
\toprule
\small
Component          & Layer            & Dimension     \\ \hline
Input              & Image            & (3,512,512)   \\ \hline
\emph{4×Semantic Encoder} &               &  \\ \hline
SFE $\boldsymbol{\theta}$ & Input            & (3,384,384)   \\ \hline
SE                 & Resnet18         & (512,12,12)   \\ \cline{2-3}
IBM \& IBS              & Conv2D           & (512,12,12)   \\ \hline
CAE  $f_{c}^{u}$              & CAP \& PN           & (512,$c_1$,$c_1$)   \\ \hline

\emph{Uplink Channel }    & Rician             &               \\ \hline
\emph{1×Edge Encoder} &               &  \\ \hline
ESE $\boldsymbol{\phi}$       & Input            & 4×(512,$c_1$,$c_1$)   \\ \hline
BE                 & Upsampling       & 4×(512,16,16) \\
                   & Reshape          & (512,32,32)   \\
                   & Conv2D           & (512,32,32)   \\
                   & Adaptive Pooling & (512,8,8)     \\ \cline{2-3}
IBM \& IBS                & Conv2D           & (512,8,8)     \\ \hline
ECE   $f_{c}^{d}$     & PN           & (512,8,8)    \\ \hline
\emph{Downlink Channel}   & Rayleigh             &               \\ \hline
\emph{3×Decoder} &               &  \\ \hline
User Decoder $\boldsymbol{\psi}$      & Input            & (512,8,8)     \\ \cline{2-3}
BD                 & Upsampling       & (512,64,64)   \\
                   & DeepLabHead      & ($c_2$,64,64)     \\
                   & Upsampling       & ($c_2$,512,512)   \\ \hline
\emph{Auxiliary Network}  $\boldsymbol{\lambda}$ &               &  \\ \hline
AE                 & Upsampling       & 4×(512,16,16) \\
                   & Reshape          & (512,32,32)   \\
                   & Conv2D           & (512,32,32)   \\
                   & Adaptive Pooling & (512,8,8)     \\ \cline{2-3}
3×AD                 & Upsampling       & (512,64,64)   \\
                   & DeepLabHead      & ($c_2$,64,64)     \\
                   & Upsampling       & ($c_2$,512,512)   \\ \hline

\bottomrule
\end{tabular}%
}
\label{tab_network_p}
\end{table}

\begin{table}[!t] 
\centering
\caption{Architecture of DeepJSCC Benchmark}
\resizebox{\columnwidth}{!}{%
\begin{tabular}{lll} 
\toprule
\small
Component          & Layer            & Dimension     \\ \hline
Input              & Image            & (3,512,512)   \\ \hline
\emph{4×Semantic Encoder} &  \emph{Same as the proposed}         &  \\ \hline
\emph{Uplink Channel }    & Rician             &               \\ \hline 
\emph{Auxiliary Decoder} &               &  \\ \hline
4×Decoder          & Input            & (512,$c_1$,$c_1$)     \\ \hline 
BD                 & Upsampling       & (512,64,64)   \\
                   & DeepLabHead      & (3,64,64)     \\
                   & Upsampling       & (3,384,384)   \\ \hline 
\bottomrule
\end{tabular}%
}
\label{tab_network_J}
\end{table}

\begin{table}[!t] 
\centering
\caption{Architecture of Decoder in Unicast Benchmark}
\resizebox{\columnwidth}{!}{%
\begin{tabular}{lll} 
\toprule
\small
Component          & Layer            & Dimension     \\ \hline
Input              & Features         & 4×(512,$c_1$,$c_1$)   \\ \hline
\emph{3×Edge Encoder} &               &  \\ \hline
ESE $\boldsymbol{\phi}$       & Input            & 4×(512,$c_1$,$c_1$)   \\ \hline
BE                 & Upsampling       & 4×(512,16,16) \\
                   & Reshape          & (512,32,32)   \\
                   & Conv2D           & (512,32,32)   \\
                   & Adaptive Pooling & (512,$c_3$,$c_3$)     \\ \cline{2-3}
IBM \& IBS                & Conv2D           & (512,$c_3$,$c_3$)     \\ \hline
ECE   $f_{c}^{d}$    & Power Normalization           & (512,$c_3$,$c_3$)    \\ \hline
\emph{Downlink Channel}   & Rayleigh             &               \\ \hline
\emph{3×Decoder} &   \emph{Same as the proposed}            &  \\ \hline
\bottomrule
\end{tabular}%
}
\label{tab_network_U}
\end{table}

\subsubsection{Benchmarks} \label{bench}
We compare the proposed edge sensing method with the following representative benchmark methods:
\begin{itemize}
\item \textbf{Ideal Transmission:} We consider a performance upper bound with ideal transmission without channel noise in the downlink. The NN structures, bandwidth allocations, and training methods are the same as the proposed scheme.
\item \textbf{DeepJSCC \cite{deepJSCC}:} JSCC utilizes NNs to perform the feature extraction and channel coding for reconstructing the original source information at the receiver side. For a fair comparison, we apply DeepJSCC to train the upstream semantic encoder. The NN structure and training procedure of the downstream edge encoder and user decoders follow that of the proposed approach. 
\item \textbf{Unicast \cite{unicast}:} The ES transmits separate features to different downstream tasks through dedicated downlink channels. We employ three edge encoders on the ES to perform feature encoding and transmission for the three tasks. The NN structure and the training of upstream semantic encoders follow that of the proposed scheme. 
\end{itemize}

The NN architecture of the DeepJSCC based benchmark is shown in Table \ref{tab_network_J}. Notice that it includes a semantic decoder for reconstructing the source data at the ES. Besides, the NN architecture of the Unicast benchmark method is shown in Table \ref{tab_network_U}. Notice that we use 3 separate edge encoders for different downstream tasks, where they share the total downlink communication bandwidth. Here, we set the variable output dimension ``$c_3$" of the edge encoder as 6, 4, and 4 for the semantic segmentation, human parts segmentation, and saliency detection, respectively. 
This setting results in a total downstream bandwidth of $512\times(6\times6+4\times4+4\times4)=512\times68$, which is slightly larger than the $512\times(8\times8)=512\times64$ used in the proposed method, so the comparison is fair with this setting. 

\begin{figure*}[!t]
\centering
\includegraphics[width=6.5in]{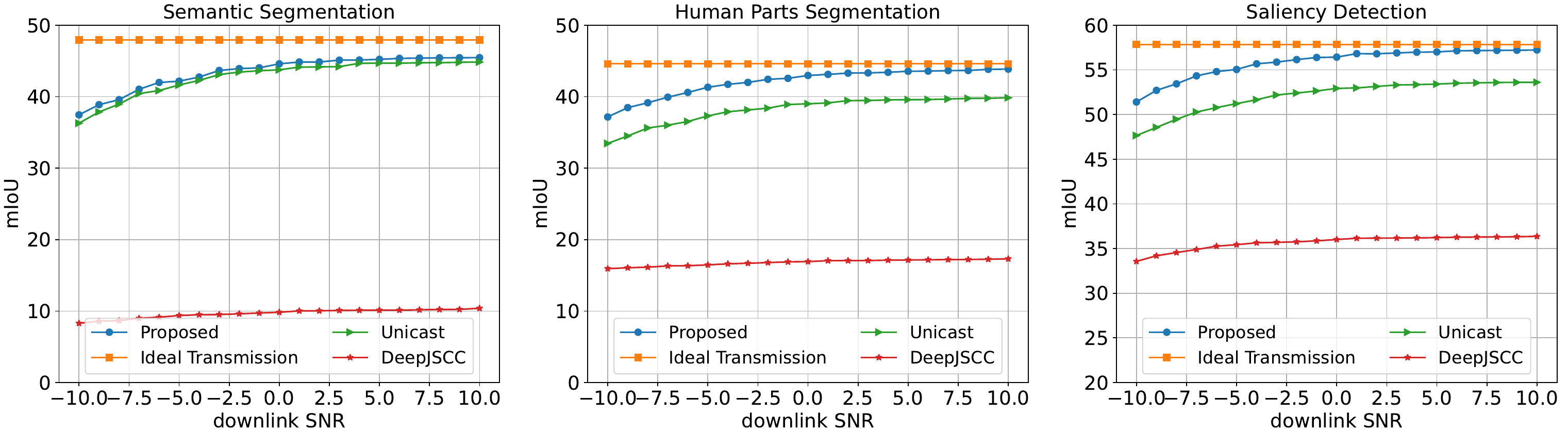}
\caption{Multi-task performance comparisons under different downlink channel SNRs (the average uplink channel SNR is 0dB).}
\label{fig_5}
\end{figure*}

\begin{figure*}[!t]
\centering
\includegraphics[width=6in]{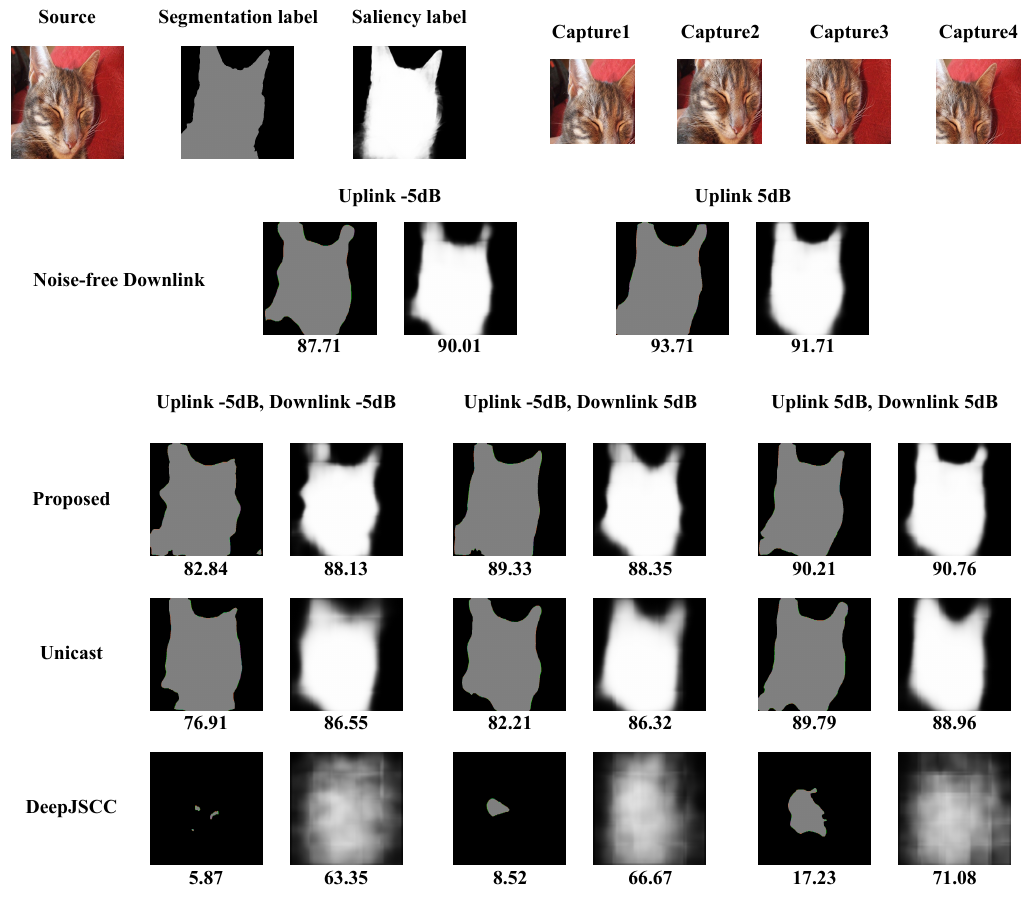}
\caption{Visualization results of semantic segmentation and saliency detection for a picture of a cat in the test set under different channel conditions. The numbers under the semantic segmentation predictions in the figure indicate the Intersection over Union (IoU) predicted by the model as “cat”, while the numbers under the saliency detection predictions indicate the IoU predicted by the model with a threshold greater than 0.5.}
\label{fig_8}
\end{figure*}

\begin{figure*}[!t]
\centering
\includegraphics[width=6.5in]{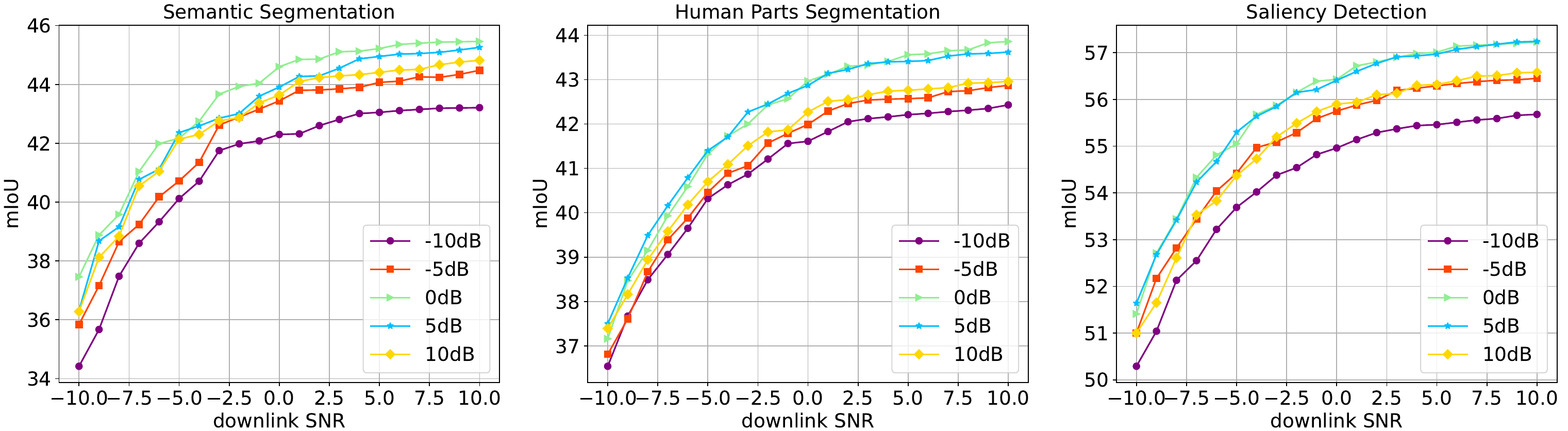}
\caption{Performance of the proposed scheme under different uplink and downlink channel SNRs. The legend in the figure denotes different uplink SNRs. }
\label{fig_6}
\end{figure*}

\subsection{Performance Comparisons} \label{PC}
We first compare in Fig. \ref{fig_5} the performance of the proposed edge sensing method with the considered benchmarks under different downlink channel conditions. Notice that the average uplink SNR is 0dB, while the instantaneous SNRs of testing samples are within $0\pm5$dB. The horizontal coordinate corresponding to each point in the figure is the average SNR of all test samples.  
The performance of the three sensing tasks is evaluated based on the \emph{mean Intersection over Union} (mIoU) \cite{M3VIT}, which is defined by the overlap between the predicted area and the ground truth, divided by the total area covered by the union of the two. As expected, the performance of all the methods improves with the SNR.
Our proposed method achieves an average performance of 43.45, 42.08, and 55.84 for semantic segmentation, human parts segmentation, and saliency detection tasks, respectively. Compared to the ideal downstream transmission, it achieves a performance degradation of 9.76\%, 5.18\%, and 3.47\%, respectively.
Compared to the Unicast method, the proposed scheme outperforms by 1.38\%, 10.24\%, and 7.19\%, respectively. 
The performance advantage is much more significant compared to the DeepJSCC method, i.e., reaching 76.3\%, 59.6\%, and 36.1\% for the three tasks.

\begin{figure*}[!t]
\centering
\includegraphics[width=6.5in]{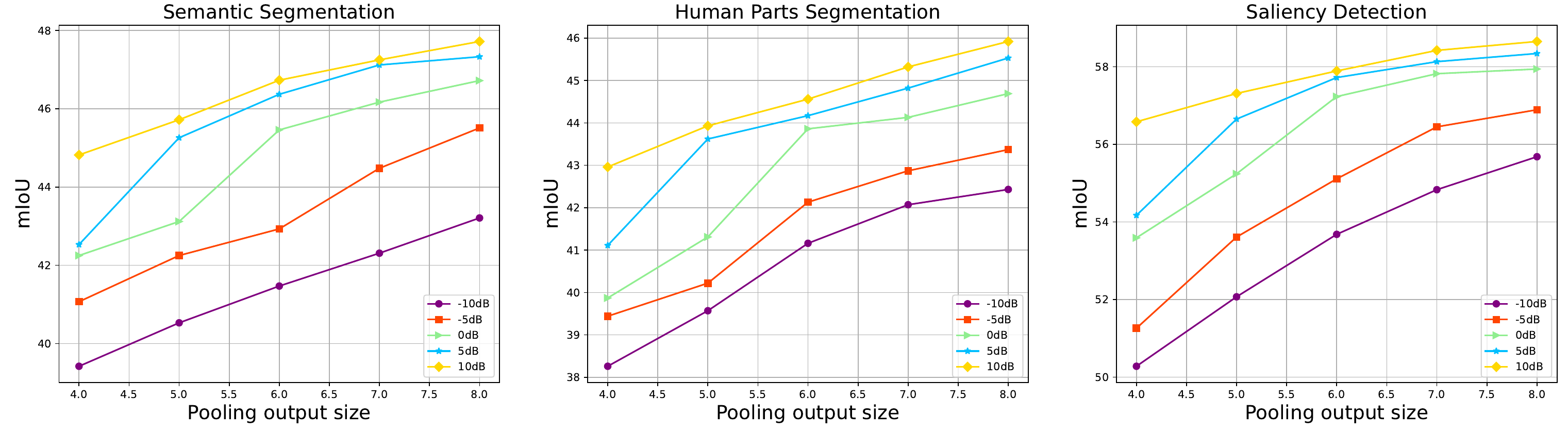}
\caption{Performance of the proposed scheme under different pooling output sizes of the CAP. The legend in the figure denotes different uplink SNRs. }
\label{fig_pooling}
\end{figure*}

Fig. \ref{fig_8} provides an example to visualize the inference performance of different methods. 
The original image, the ground-truth labels of segmentation and saliency detection tasks, and the multi-view pictures taken by four SDs are shown at the top of Fig. \ref{fig_8}. We then compare the inference results of different methods under several combinations of uplink and downlink SNRs. Meanwhile, we also considers transmission under noiseless downlink condition as a performance upper bound of the proposed scheme. The corresponding Intersection over Union (IoU) value is presented at the bottom of each inference result. A close observation shows that the proposed method accurately locates and outlines the object (cat) under different SNR conditions, which outperforms the Unicast method, especially under low SNR conditions. Notably, the DeepJSCC method suffers the most significant degradation, failing to detect the object under all channel conditions.

The result demonstrates the performance advantage of the joint design of uplink feature extraction and downlink information broadcast. Compared to the DeepJSCC scheme, the result shows the critical importance of applying task-oriented feature extraction in the uplink. Besides, compared to the Unicast scheme, the result shows that applying the proposed semantic broadcast technique can further improve the task performance under stringent bandwidth constraints.

\subsection{Accuracy-Bandwidth Tradeoffs}
In this subsection, we examine the accuracy-bandwidth tradeoffs achieved by the proposed edge sensing scheme. Recall that we have introduced a CAP layer at the encoder of SDs, which reduces the output dimension at a higher SNR.  Fig. \ref{fig_6} presents the performance of semantic segmentation, human parts segmentation, and saliency detection tasks, respectively. Table \ref{tab_SNR_CR} illustrates the CRs (defined as the ratio between the output dimension and the input dimension) achieved with the CAP module and the corresponding performance degradation compared to the max pooling output size (i.e., $l_p=8$ for all uplink channel SNRs). 
From Fig. \ref{fig_6} and Table \ref{tab_SNR_CR}, we observe that, with the adaptive pooling method, the performance variation of all tasks is less than 5\%. Meanwhile, compared to using the largest output dimension at -10dB, 
application of the CAP module reduces the uplink data volume by 43.75\% at 0dB, and by 60.94\% and 75\% at 5dB and 10dB, respectively. 
This shows that the proposed adaptive pooling method can achieve significant uplink bandwidth conservation with minor performance degradation compared to setting a fixed and maximum pooling output size. 

Fig. \ref{fig_pooling} shows the impact of pooling output size (i.e., $c_1$ in Table I) on the mIoU performance of downstream tasks. As expected, the mIoU of all three tasks grows with the pooling output size. 
Besides, the mIoU performance improves with a higher uplink SNR. Based on the results, we can select a proper pooling output size to meet the inference requirement, while minimizing the occupied communication bandwidth. For instance, when the uplink SNR = -5 dB, and we require that the mIoU$\geq 43$ for all the three tasks, we can select a pooling output size $7.0$ to minimize the communication bandwidth.

\begin{table}[!t] 
\centering
\caption{CR and Performance Degradation with CAP}
\resizebox{\columnwidth}{!}{%
\begin{tabular}{llllll} 
\toprule
SNR &   -10dB     & -5dB      & 0dB     & 5dB & 10dB   \\ \hline
CR   & 0.0741    & 0.0567   & 0.0417  & 0.0289  & 0.0185  \\ \hline
Reduced data & 0    & 23.44\%   & 43.75\%  & 60.94\%  & 75\%  \\ \hline
Degradation Seg  & 0    & 1.13\%   & 1.98\%   & 2.41\%  & 2.73\%  \\ \hline
Degradation Hum  & 0    & 0.32\%   & 2.02\%  & 2.49\%  & 4.63\%  \\ \hline
Degradation Sal  & 0    & 0.12\%   & 1.11\%  & 2.35\%  & 3.88\%  \\ \hline
\bottomrule
\end{tabular}%
}
\label{tab_SNR_CR}
\end{table}

\begin{figure*}[!t]
\centering
\includegraphics[width=6.5in]{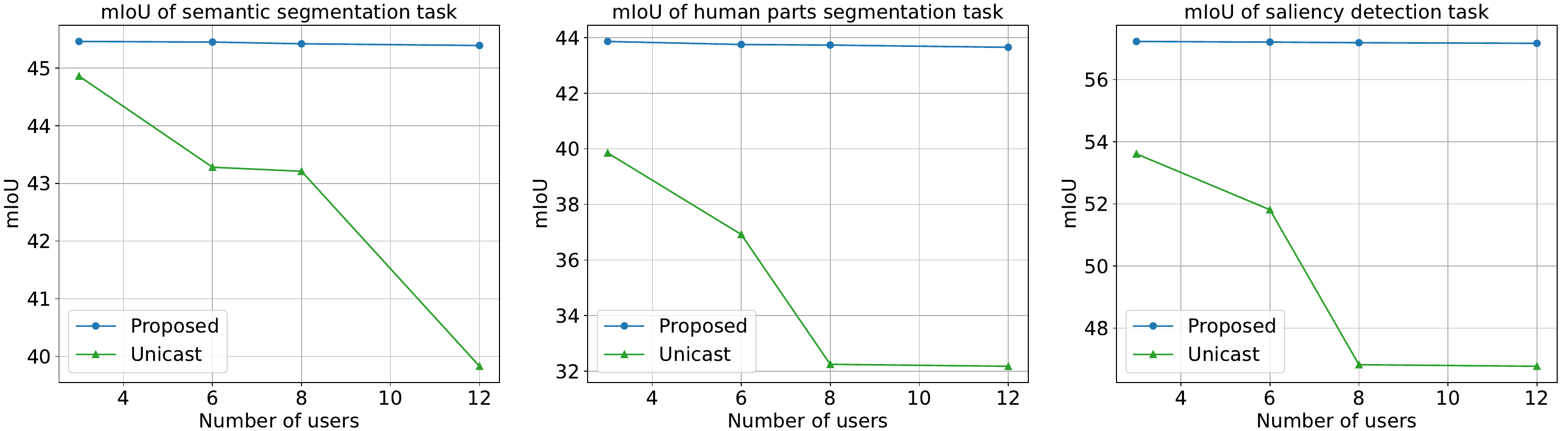}
\caption{Multi-task performance with different number of users.}
\label{fig_9}
\end{figure*}

\subsection{Scalability Performance}
In this subsection, we evaluate the scalability of the proposed scheme when the number of downstream users and tasks increases. Without loss of generality, the uplink channel SNR $\bar{\Upsilon}^u$ is 0dB and the downlink channel SNR $\bar{\Upsilon}^d$ is 10dB for each user. Besides, the bandwidth of the proposed broadcast scheme is $512\times8\times8=512\times64$. For the compared Unicast method, the total bandwidth is $512\times68$, which will be equally split to the multiple downstream users. 

We first test the system performance when the number of users increases. Specifically, we consider the number of downstream users $M \in\{3,6,8,12\}$, train the downstream edge encoder and users' corresponding decoders, and examine the variation of inference accuracy and latency. For the proposed broadcast method, the downstream users receive semantic features (generated by a common edge encoder) of the same dimension ($512\times8\times8=512\times64$), but each user receives the features with their respective downlink noise. For the contrasting Unicast method, a separate bandwidth allocation is required for each user. Specifically, for the 6-user scenario, the number of users performing semantic segmentation, human parts segmentation, and saliency detection task is set equal as $2$. The total downstream bandwidth of the Unicast method is $512\times(4\times4\times2+3\times3\times2+3\times3\times2)=512\times68$. For the 8-user scenario, the number of users for the three tasks are 3, 3, and 2, respectively. The total downstream bandwidth is $512\times(4\times4\times3+2\times2\times3+2\times2\times2)=512\times68$. For the 12-user scenario, 4 users perform each task, where the total downstream bandwidth is $512\times(3\times3\times4+2\times2\times4+2\times2\times4)=512\times68$. Therefore, for different number of users, we use similar downlink communication bandwidth for fair comparisons. 

\begin{table}[!t] 
\centering
\caption{Inference Time Comparison for Different Number of Users (in milliseconds)}
\resizebox{\columnwidth}{!}{%
\begin{tabular}{llllll} 
\toprule
Method &  Semantic Encoder     & Edge Encoder      & User Decoder   \\ \hline
Proposed   & 75.5    & 4.1   & 26.2      \\ \hline
Unicast 3-user    & 75.5    & 12.3   & 26.2      \\ \hline
Unicast 6-user   & 75.5    & 24.5   & 26.2       \\ \hline
Unicast 8-user   & 75.5    & 32.8   & 26.2       \\ \hline
Unicast 12-user   & 75.5    & 49.1   & 26.2       \\ \hline
\bottomrule
\end{tabular}%
}
\label{tab_inference}
\end{table}

\begin{table}[!t]
    \centering
    \caption{{MIoU Performance Comparison for Different Number of Tasks}}
    \begin{tabular}{cccc}
    \toprule
        Method & 1 Task & 2 Tasks & 3 Tasks \\ \hline
        Proposed & Seg: 45.48 & \makecell{Seg: \textbf{45.46}\\ Sal: \textbf{57.17}} & \makecell{Seg: \textbf{45.46}\\ Sal: \textbf{57.16}\\ Hum: \textbf{43.65}} \\ \hline
        Unicast & Seg: 45.48 & \makecell{Seg: 45.07\\ Sal: 54.43}  & \makecell{Seg: 43.86\\Sal: 53.61\\Hum: 39.85} \\ \hline
    \bottomrule
    \end{tabular}
    \label{tab:inference2}
\end{table}

\begin{table}[!t]
\centering
\caption{ Inference Time Comparison (Edge Encoder) for Different Number of Tasks (in milliseconds)}
\small
\begin{tabular}{llllll} 
\toprule
 &  1 Task     & 2 Tasks      & 3 Tasks   \\ \hline
Proposed   & 4.1   & 4.1   & 4.1      \\ \hline
Unicast    & 4.1    & 8.2   & 12.3      \\ \hline
\bottomrule
\label{tab:inference3}
\end{tabular}%
\end{table}

We first evaluate the average user performance for the 3 tasks. As shown in Fig. \ref{fig_9}, all the three tasks of the proposed scheme degrades less than 1\% when the number of users increases from 3 to 12.  
In contrast, the performance of the Unicast method decreases by 9.6\%, 18.8\%, and 12.4\%, respectively, for the three tasks.  
The performance degradation of the Unicast scheme is due to the smaller output feature dimension allocated to each downstream user when $M$ increases. For instance, each user of the saliency detection task is allocated with $512\times 2 \times 2$ communication bandwidth when $M =12$. However, the proposed scheme utilizes a constant $512 \times 8 \times 8$ communication bandwidth thanks to the broadcast nature of the downlink transmission design.

\begin{figure*}
\begin{align} 
&\quad  -\sum_{m=1}^{M} \alpha _m I_{\boldsymbol{\theta} }(Y_{m};W)+\beta \cdot \sum_{n=1}^{N}I_{\boldsymbol{\theta} _n}(X_n;Z_n) \notag \\
&= - \sum_{m=1}^{M} \alpha _m \int p(\boldsymbol{y}_{m}|\boldsymbol{w}) p(\boldsymbol{w}) \log \frac{p(\boldsymbol{y}_{m}|\boldsymbol{w})}{p(\boldsymbol{y}_{m})} d\boldsymbol{y} d\boldsymbol{w} 
+  \beta \cdot \sum_{n=1}^{N} \int p_{\boldsymbol{\theta} _n}(\boldsymbol{z}_{n}|\boldsymbol{x}_{n}) p(\boldsymbol{x}_{n}) \log \frac{p_{\boldsymbol{\theta} _n}(\boldsymbol{z}_{n}|\boldsymbol{x}_{n})}{p(\boldsymbol{z}_{n})} d\boldsymbol{x} d\boldsymbol{z} \notag \\
&= \underbrace{ - \sum_{m=1}^{M} \alpha _m \int p(\boldsymbol{y}_{m}|\boldsymbol{w}) p(\boldsymbol{w}) \log p(\boldsymbol{y}_{m}|\boldsymbol{w}) d\boldsymbol{y} d\boldsymbol{w} +  \beta \cdot \sum_{n=1}^{N} \int p_{\boldsymbol{\theta} _n}(\boldsymbol{z}_{n}|\boldsymbol{x}_{n}) p(\boldsymbol{x}_{n}) \log \frac{p_{\boldsymbol{\theta} _n}(\boldsymbol{z}_{n}|\boldsymbol{x}_{n})}{p(\boldsymbol{z}_{n})} d\boldsymbol{x} d\boldsymbol{z} }_{\mathcal{L}_{MIB}(\boldsymbol{\theta})} - H(Y) \label{IB_theta}  
\end{align}
\begin{align}
&\quad - \sum_{m=1}^{M} \alpha _m \int p(\boldsymbol{y}_{m}|\boldsymbol{w}) p(\boldsymbol{w}) \log p(\boldsymbol{y}_{m}|\boldsymbol{w}) d\boldsymbol{y} d\boldsymbol{w} \notag \\
&= - \sum_{m=1}^{M} \alpha _m \int p(\boldsymbol{y}_{m}|\boldsymbol{w}) p(\boldsymbol{w}) \log q_{\boldsymbol{\lambda}}(\boldsymbol{y}_{m}|\boldsymbol{w}) d\boldsymbol{y} d\boldsymbol{w} - \sum_{m=1}^{M} \alpha _m \underbrace{\int p(\boldsymbol{y}_{m}|\boldsymbol{w}) p(\boldsymbol{w}) \log \frac{p(\boldsymbol{y}_{m}|\boldsymbol{w})}{q_{\boldsymbol{\lambda}}(\boldsymbol{y}_{m}|\boldsymbol{w})} d\boldsymbol{y} d\boldsymbol{w}}_{=D_{KL}(p(\boldsymbol{y}_{m}|\boldsymbol{w}) \parallel q_{\boldsymbol{\lambda}}(\boldsymbol{y}_{m}|\boldsymbol{w})) \ge 0} \label{IB_theta_left} 
\end{align}
\begin{align}
&\quad  \sum_{n=1}^{N} \int p_{\boldsymbol{\theta} _n}(\boldsymbol{z}_{n}|\boldsymbol{x}_{n}) p(\boldsymbol{x}_{n}) \log \frac{p_{\boldsymbol{\theta} _n}(\boldsymbol{z}_{n}|\boldsymbol{x}_{n})}{p(\boldsymbol{z}_{n})} d\boldsymbol{x} d\boldsymbol{z} \notag \\
&=  \sum_{n=1}^{N} \int p_{\boldsymbol{\theta} _n}(\boldsymbol{z}_{n}|\boldsymbol{x}_{n}) p(\boldsymbol{x}_{n}) \log \frac{p_{\boldsymbol{\theta} _n}(\boldsymbol{z}_{n}|\boldsymbol{x}_{n})}{q(\boldsymbol{z}_{n})} d\boldsymbol{x} d\boldsymbol{z} -  \sum_{n=1}^{N} \underbrace{\int p_{\boldsymbol{\theta} _n}(\boldsymbol{z}_{n}|\boldsymbol{x}_{n}) p(\boldsymbol{x}_{n}) \log \frac{p(\boldsymbol{z}_{n})}{q(\boldsymbol{z}_{n})} d\boldsymbol{x} d\boldsymbol{z}}_{=D_{KL}(p(\boldsymbol{z}_{n} \parallel q(\boldsymbol{z}_{n})) \ge 0} \label{IB_theta_right} 
\end{align}
\begin{align}
\mathcal{L}_{MVIB}(\boldsymbol{\theta}, \boldsymbol{\lambda}) &= - \sum_{m=1}^{M} \alpha _m \int p(\boldsymbol{y}_{m}|\boldsymbol{w}) p(\boldsymbol{w}) \log q_{\boldsymbol{\lambda}}(\boldsymbol{y}_{m}|\boldsymbol{w}) d\boldsymbol{y} d\boldsymbol{w} +  \beta \cdot \sum_{n=1}^{N} \int p_{\boldsymbol{\theta} _n}(\boldsymbol{z}_{n}|\boldsymbol{x}_{n}) p(\boldsymbol{x}_{n}) \log \frac{p_{\boldsymbol{\theta} _n}(\boldsymbol{z}_{n}|\boldsymbol{x}_{n})}{q(\boldsymbol{z}_{n})} d\boldsymbol{x} d\boldsymbol{z} \label{IB_theta_lambda} 
\end{align}
{\noindent} \rule[-10pt]{18cm}{0.05em}
\end{figure*}

In Table \ref{tab_inference}, we then evaluate the inference time (in milliseconds) of the semantic encoder deployed on the edge SD, the edge encoder on the ES, and the decoder on the user in the above multi-user scenario. 
For both the proposed scheme and the Unicast method, the computational complexity of the semantic encoder and decoder is irrelevant to the user number. Therefore, the inference delay of the SD and user keeps constant when $M$ varies.
Meanwhile, the proposed scheme employs only one edge encoder, where the inference delay does not vary with $M$ as well. In contrast, the edge inference time of the Unicast method increases almost linearly with the number of downstream users, e.g., from 12.3 milliseconds for 3 users to 49.1 milliseconds for 12 users. This is because that the Unicast method employs $M$ edge encoders, one for each user, leading to $M$ times inference computational complexity. To achieve the same inference time, in practice, the Unicast scheme can implement $M$ parallel edge encoders, one for each task/user. This, however, also increases the computational power required at the ES by $M$ times compared to the proposed scheme. 

We also evaluate the performance and inference time when the number of tasks increases. Specifically, we consider initially the network serving only one user, and then gradually adding users with new tasks to the network. Notice that in the 1 task scenario, the bandwidth allocation of the Unicast is the same as the proposed scheme. In the 2 tasks scenario, the Unicast method allocates $512\times 6\times 6$ transmission feature size to each task. In the 3 tasks, the allocation is the same as depicted in Section \ref{bench}.
It shows that, as we add new tasks to a network, the performance of existing tasks remains almost unchanged for the proposed scheme. In comparison, the Unicast scheme achieves worse inference performance than the proposed scheme for the two newly added tasks. Besides, Table \ref{tab:inference3} shows the inference time (in milliseconds) of the edge encoder in the ES for different numbers of downstream tasks. As shown in Table \ref{tab:inference3}, for the Unicast method, the ES has to encode features separately for each task, so the inference time grows linearly with the number of tasks. But for the proposed broadcast method, the inference time is independent of the number of tasks, because the ES only needs to encode and broadcast once. The above results have verified the scalability of the proposed method when the task number increases.

Overall, the performance and inference latency of the proposed method are not affected by the number of users or tasks, showing strong scalability when the network size increases. 

\begin{figure*}[htbp]
\begin{align} 
&\quad  -\sum_{m=1}^{M} \gamma _m I_{\boldsymbol{\phi} }(Y_{m};U_{m})+  \eta \cdot I_{\boldsymbol{\phi}}(S;W) \notag \\
&= - \sum_{m=1}^{M} \gamma _m \int p(\boldsymbol{y}_{m}|\boldsymbol{v}_{m}) p(\boldsymbol{v}_{m}) \log \frac{p(\boldsymbol{y}_{m}|\boldsymbol{v}_{m})}{p(\boldsymbol{y}_{m})} d\boldsymbol{y} d\boldsymbol{v} 
+  \eta \cdot \int p_{\boldsymbol{\phi}}(\boldsymbol{s}|\boldsymbol{w}) p(\boldsymbol{w}) \log \frac{p_{\boldsymbol{\phi} }(\boldsymbol{s}|\boldsymbol{w})}{p(\boldsymbol{w})} d\boldsymbol{w} d\boldsymbol{s} \notag \\
&= \underbrace{ - \sum_{m=1}^{M} \gamma _m \int  p(\boldsymbol{y}_{m}|\boldsymbol{v}_{m}) p(\boldsymbol{v}_{m}) \log p(\boldsymbol{y}_{m}|\boldsymbol{v}_{m}) d\boldsymbol{y} d\boldsymbol{v} +   \eta \cdot \int p_{\boldsymbol{\phi}}(\boldsymbol{s}|\boldsymbol{w}) p(\boldsymbol{w}) \log \frac{p_{\boldsymbol{\phi} }(\boldsymbol{s}|\boldsymbol{w})}{p(\boldsymbol{w})} d\boldsymbol{w} d\boldsymbol{s} }_{\mathcal{L}_{MIB}(\boldsymbol{\phi}, \boldsymbol{\psi})} - H(Y) \label{IB_phi} 
\end{align}
\begin{align}
&\quad - \sum_{m=1}^{M} \gamma _m \int  p(\boldsymbol{y}_{m}|\boldsymbol{v}_{m}) p(\boldsymbol{v}_{m}) \log p(\boldsymbol{y}_{m}|\boldsymbol{v}_{m}) d\boldsymbol{y} d\boldsymbol{v}  \notag  \\
&= - \sum_{m=1}^{M} \gamma _m \int  p(\boldsymbol{y}_{m}|\boldsymbol{v}_{m}) p(\boldsymbol{v}_{m}) \log q_{\boldsymbol{\psi}_{m}}(\boldsymbol{y}_{m}|\boldsymbol{v}_{m}) d\boldsymbol{y} d\boldsymbol{v} - \sum_{m=1}^{M} \gamma _m  \underbrace{ \int  p(\boldsymbol{y}_{m}|\boldsymbol{v}_{m}) p(\boldsymbol{v}_{m}) \log \frac{p(\boldsymbol{y}_{m}|\boldsymbol{v}_{m})}{q_{\boldsymbol{\psi}_{m}}(\boldsymbol{y}_{m}|\boldsymbol{v}_{m})} d\boldsymbol{y} d\boldsymbol{v} }_{ = D_{KL}(p(\boldsymbol{y}_{m}|\boldsymbol{v}_{m}) \parallel q_{\boldsymbol{\psi}_{m}}(\boldsymbol{y}_{m}|\boldsymbol{v}_{m})) \ge 0 } \label{IB_phi_l}
\end{align}
\begin{align}
  \int p_{\boldsymbol{\phi}}(\boldsymbol{s}|\boldsymbol{w}) p(\boldsymbol{w}) \log \frac{p_{\boldsymbol{\phi} }(\boldsymbol{s}|\boldsymbol{w})}{p(\boldsymbol{s})} d\boldsymbol{w} d\boldsymbol{s}  
=   \int p_{\boldsymbol{\phi}}(\boldsymbol{s}|\boldsymbol{w}) p(\boldsymbol{w}) \log \frac{p_{\boldsymbol{\phi} }(\boldsymbol{s}|\boldsymbol{w})}{q(\boldsymbol{s})} d\boldsymbol{w} d\boldsymbol{s}  - \underbrace{\int p_{\boldsymbol{\phi}}(\boldsymbol{s}|\boldsymbol{w}) p(\boldsymbol{w})  \log \frac{p(\boldsymbol{s})}{q(\boldsymbol{s})} d\boldsymbol{w} d\boldsymbol{s} }_{=D_{KL}(p(\boldsymbol{s}) \parallel q(\boldsymbol{s})) \ge 0} \label{IB_phi_r}
\end{align}
\begin{align} \label{L_VIB_D}
\mathcal{L}_{VMIB}(\boldsymbol{\phi}, \boldsymbol{\psi}) = - \sum_{m=1}^{M} \gamma _m \int  p(\boldsymbol{y}_{m}|\boldsymbol{v}_{m}) p(\boldsymbol{v}_{m}) \log q_{\boldsymbol{\psi}_{m}}(\boldsymbol{y}_{m}|\boldsymbol{v}_{m}) d\boldsymbol{y} d\boldsymbol{v} +  \eta \cdot \int p_{\boldsymbol{\phi}}(\boldsymbol{s}|\boldsymbol{w}) p(\boldsymbol{w}) \log \frac{p_{\boldsymbol{\phi} }(\boldsymbol{s}|\boldsymbol{w})}{q(\boldsymbol{s})} d\boldsymbol{w} d\boldsymbol{s}
\end{align}
{\noindent} \rule[-10pt]{18cm}{0.05em}
\end{figure*}

\section{Conclusions and Future Work} \label{clu}
In this paper, we proposed a scalable task-oriented edge sensing framework for multi-task execution. In the proposed framework, the ES aggregates the processing workload for correlated sensing tasks, performing joint semantic
information gathering, and broadcast in a task-oriented fashion.
We extend the well-known IB theory to a multi-task scenario to jointly optimize the performance of information gathering and broadcast.
To adapt various uplink and downlink channel conditions, we develop an efficient two-step training procedure of NN-based codecs used in the framework. Experiment results demonstrate that, compared to the DeepJSCC method, the proposed method can extract task-relevant features effectively, and achieves an average lead of 57.3\% in the 3-task scenario. Besides, with the proposed CAP, the uplink data volume reduces 60.94\% and 75\% at 5dB and 10dB SNR, while the performance degradation is less than 5\%. 
Compared to the Unicast method, the proposed scheme outperforms by an average performance of 6.3\%. Meanwhile, the experiments highlight that with the same downstream bandwidth, the proposed method achieves less than 1.2\% performance degradation and constant computation delay when the number of users increases from 3 users to 12 users. 

We conclude the paper with in this paper, we design the CAP method to realize the trade-off between performance and bandwidth. Because of the discrete set of output dimensions available, in some cases, the inference accuracy achieved at a higher SNR is made lower than that achieved at a lower SNR. It is therefore interesting and important to design a more refined control mechanism to adjust the output feature dimension so that the accuracy can be made non-decreasing with the SNR. A promising solution is to use dynamic neural networks \cite{shao1} that can adjust the active dimension of the output feature in the unit of a single neuron based on instantaneous SNR.

\appendices

\section{Derivation of VMIB in (\ref{L_VIB1})} \label{P_VI}
We show in (\ref{IB_theta}) the detailed computation of the objective function of (\ref{IB_fe}). Notice that the information entropy $H(Y)$ in (\ref{IB_theta}) is a constant. Therefore, by neglecting the constant term $H(Y)$ and converting the integral into a mathematical expectation form, we can equivalently minimize (\ref{L_IB}) to solve (\ref{IB_fe}). 
Then, for the left and right terms in (\ref{IB_theta}), we use the two variational distributions $q_{\boldsymbol{\lambda} } (\boldsymbol{y}_{m}|\boldsymbol{w})$  and $q(\boldsymbol{z}_n)$ to approximate $p (\boldsymbol{y}_{m}|\boldsymbol{w})$ and $p(\boldsymbol{z}_n)$, and expand the two terms in (\ref{IB_theta_left}) and (\ref{IB_theta_right}), respectively. Notice that the variational distribution $q_{\boldsymbol{\lambda} } (\boldsymbol{y}_{m}|\boldsymbol{w})$ is generated by the auxiliary NN $\boldsymbol{\lambda}$. 

From (\ref{IB_theta_left}) and (\ref{IB_theta_right}), we see that the terms on the right-hand side of them are non-negative KL divergence terms. Hence, by combining the terms on the left-hand side of (\ref{IB_theta_left}) and (\ref{IB_theta_right}), we obtain a variational expression in (\ref{IB_theta_lambda}), which is an upper bound of (\ref{IB_theta}). By converting the integral of (\ref{IB_theta_lambda})  into a mathematical expectation form, we obtain the VMIB expression in (\ref{L_VIB1}).

\section{Derivation of VMIB in (\ref{L_VIB_de})}  \label{P_VI_D}
We show the detailed computation of the objective function (\ref{IB_de}) in (\ref{IB_phi}). Since the information entropy $H(Y)$ is a constant, we can equivalently minimize (\ref{L_IB_de}) to solve (\ref{IB_de}) by neglecting the constant term $H(Y)$ and converting the integral into a mathematical expectation form. 
We then use two variational distributions $q_{\boldsymbol{\psi}_m}(\boldsymbol{y}_m|\boldsymbol{v}_m)$ and $q(\boldsymbol{s})$ to approximate $p(\boldsymbol{y}_m|\boldsymbol{v}_{m})$ and $p(\boldsymbol{s})$, as shown in (\ref{IB_phi_l}) and (\ref{IB_phi_r}). The distribution $q_{\boldsymbol{\psi}_m}(\boldsymbol{y}_m|\boldsymbol{v}_m)$ is generated by the $m$th downstream decoder. 
Notice that the terms on the right-hand side of (\ref{IB_phi_l}) and (\ref{IB_phi_r}) are non-negative KL divergence terms. Hence, by combining the terms on the left-hand side of (\ref{IB_phi_l}) and (\ref{IB_phi_r}), we obtain a variational expression in (\ref{L_VIB_D}), which is an upper bound of (\ref{IB_phi}). Finally, by converting the integral of (\ref{L_VIB_D}) into a mathematical expectation form, we obtain the VMIB expression in (\ref{L_VIB_de}).



\ifCLASSOPTIONcaptionsoff
  \newpage
\fi

\end{document}